\documentclass[11pt,a4paper]{article} 
\usepackage{jheppub}

\usepackage{slashed}
\usepackage{youngtab}
\usepackage{amsmath}
\usepackage{genyoungtabtikz}
\usepackage{standalone}
\usepackage{booktabs}
\usepackage{subcaption}
\usepackage{pgfplots} 

\usepackage[utf8]{inputenc}
\usepackage{pgfplots}
\usepackage{ytableau}
\usetikzlibrary{tikzmark,decorations.pathreplacing}
\usepackage{float}
\usepackage{subcaption}

\newcommand{\la}{\langle}
\newcommand{\ra}{\rangle}

\newcommand{\beq}{\begin{equation}}
\newcommand{\eeq}{\end{equation}}

\newcommand{\beqa}{\begin{eqnarray}}
\newcommand{\eeqa}{\end{eqnarray}}

\newcommand{\si}{\sigma}

\newcommand{\thalf}{\tfrac{1}{2}}
\newcommand{\tthalf}{\tfrac{3}{2}}

\newcommand{\tothird}{\tfrac{1}{3}}

\newcommand\fverb{\setbox\fverbbox=\hbox\bgroup\verb}
\newcommand\fverbdo{\egroup\medskip\noindent%
            \fbox{\unhbox\fverbbox}\ }
\newcommand\fverbit{\egroup\item[\fbox{\unhbox\fverbbox}]}
\newbox\fverbbox

\newcommand{\nablaslash}{\not{\hbox{\kern-3pt $\nabla$}}}

\title{ The role of singletons  in $S^7$ compactifications }

 \author[a]{B.E.W.~Nilsson,}
\author[a]{A. Padellaro}
 \author[b]{and C.N. Pope}
\affiliation[a]{Physics\ Department\\ Chalmers University of Technology\\
SE-412 96 G\"oteborg, Sweden} 
\affiliation[b]{George P. and Cynthia Woods Mitchell Institute for Fundamental Physics and Astronomy\\
 Texas A{\&}M University\\
College Station, TX 77843, USA.}
 \emailAdd{tfebn@chalmers.se, adrian.padellaro@gmail.com, pope@physics.tamu.edu }

\abstract{ 
We derive the isometry irrep content of   squashed seven-sphere compactifications of eleven-dimensional supergravity, i.e., the left-squashed ($LS^7$) with ${\mathcal N}=1$ and right-squashed ($RS^7$) with ${\mathcal N}=0$ 
supersymmetry, in a manner completely independent of the round sphere.
Then we compare this result with the spectrum obtained by Higgsing the round sphere spectrum. This way we discover features of the spectra which makes it possible to argue
 that the only way the round  spectrum can be related by
a Higgs mechanism  to the one of $LS^7$ is if the singletons are included in the round sphere spectrum. For this to work also in the $RS^7$ case it seems that the gravitino
of the $LS^7$ spectrum must be replaced by a fermionic singleton present in the $RS^7$ spectrum.}

\keywords{Kaluza-Klein, AdS/CFT, singletons}

\toccontinuoustrue
\begin{document}

\begin{flushright}
\hfill{ MI-TH-189}
\end{flushright}

\maketitle



\section{Introduction}

The  role of singletons \cite{Dirac:1963ta} in the Freund-Rubin \cite{Freund:1980xh} compactification of eleven-dimensional supergravity on $AdS_4\times S^7$ has been unclear ever since 
the spectrum was first obtained. When deriving the spectrum \cite{Biran:1983iy, Englert:1983rn, Freedman:1983na, Sezgin:1983ik}, one  finds that the round seven-sphere has special symmetries that can be used to 
gauge away precisely the irreps $D(E_0,s)$ that  correspond to the supersingleton irrep of $SO(3,2)$, that is $D(\thalf,0)\oplus D(1,\thalf)$, see, e.g., \cite{Duff:1986hr}. 

Here we  make the natural assumption that  the $AdS_4$ supergravity theory obtained by compactifying  eleven-dimensional supergravity on the squashed seven-sphere \cite{Awada:1982pk}, either left- or right-squashed ($LS^7$ or $RS^7$),  is a Higgsed version \cite{Duff:1982ev} of the one obtained from the round sphere. As we will see below this makes it possible to argue that  the supersingleton has a key role to play in this context.  It should be emphasised that the  singletons that appear in the round seven-sphere spectrum 
are  gauged away in the bulk of $AdS_4$
by symmetries that arise precisely when the sphere is round. A common interpretation of this fact is that the supersingleton is not part of the compactification spectrum on the round sphere.
However, one could instead take the point of view that the supersingleton should be kept in the spectrum but confined to the boundary of $AdS_4$ in accordance with the standard
picture  we have of singletons, see, e.g., \cite{Flato:1999yp,Nicolai:1984gb, Duff:1986hr, Bergshoeff:1987dh}.

In order to back up  this point of view we  take advantage of a particular property of the relation between the round and left-squashed spectra demonstrated in the following sections  to show that the supersingleton must be kept in the round sphere spectrum in order for the Higgsing/deHiggsing to give the correct spectrum on the supersymmetric squashed sphere.  
As will be clear later this requires the singletons to undergo some kind of Higgsing, i.e., eating some ordinary field of the same spin so that it can become an ordinary 
scalar/spin-$\thalf$ bulk field
itself. We mention here that we, however, are not aware of any known field theory realisation of such a phenomenon in the literature and we have nothing to add to this question.
To reach this conclusion we have  constructed the  complete isometry representation content of the squashed sphere spectrum\footnote{Note that we have no new information on the eigenvalue spectrum
of the mass operators on the squashed seven-sphere some of which are known, see \cite{Duff:1986hr} and references therein.  As will be clear below, the eigenvalues are not needed to reach our conclusions about the singletons.} in a manner that is entirely  independent from the round sphere. As it turns out, using the same kind of reasoning, also the spectrum on the $RS^7$ must contain a fermionic singleton.

In Section 2 we first review the spectrum on the round sphere and then derive the complete spectrum of isometry irreps on $LS^7$ and  $RS^7$ independently of the round case. These spectra are then compared and some conclusions drawn concerning the role of singletons. In the final section we summarise our conclusions and make some additional comments. Tables of squashed sphere harmonics are collected in the Appendix.
\section{Comparison of the spectra on the round and squashed seven-spheres}

In the first subsection of this section we  give a very brief account of Freund-Rubin compactifications
of eleven-dimensional supergravity and their general features relevant for our discussion of the spectrum, see \cite{Duff:1986hr} for a more detailed discussion. 
This includes tables over unitary irreps of $SO(3,2)$ giving $E_0$ as a function of mass (fermions) or mass squared (bosons) (Table 1) and expressions for the mass operators in terms of invariant operators 
($\Delta_L$ etc) on the seven-sphere (Table 2).

Then,  in the 
second subsection, we recall the $AdS_4$ spectrum of irreps $D(E_0, s)$ of $SO(3,2)$ appearing in the 
compactification  on the round seven-sphere
including the $SO(8)$ irrep each of these transform under. This is presented in Table 3 which also contains the round sphere eigenvalues of the relevant operators.  In subsection 3  we turn to the squashed sphere spectrum and introduce a Young Tableau method by means of which we can
derive the complete spectrum of isometry irreps for each field  appearing  in the $AdS_4$ supergravity 
theory here choosing the orientation
that leads to an ${\mathcal N}=1$ supersymmetric theory. This case is referred to as the left-squashed ($LS^7$) case while its 
orientation-flipped (or skew-whiffed) right-squashed ($RS^7$) cousin has no supersymmetries. This latter case will also play a role in 
this paper and we will  comment on it both here in the final subsection and in  the Conclusions. The fourth and final subsection is  devoted to a comparison between the 
Higgsed/deHiggsed version of the 
spectrum on the round sphere and the spectrum obtained directly on the squashed sphere in  subsection 3.
Tables of squashed sphere harmonics for all the relevant operators are collected in the Appendix\footnote{Note that the irrep spectrum, contrary to the eigenvalue spectrum, is the same for the left and right squashed spheres.}. The two crucial
tables for the arguments presented here are the ones for $\Delta_L$ and ${\slashed D}_{3/2}$.

\subsection{General features of Freund-Rubin compactifications}

The theory under consideration in this paper is eleven-dimensional supergravity compactified on $AdS_4$ times either the round or squashed seven-sphere. This latter factor can have two different orientations,
left or right,  denoted  $LS^7$  and $RS^7$, having ${\mathcal N}=1$ and ${\mathcal N}=0$ supersymmetry, respectively \cite{Duff:1983ajq, Duff:1984sv}. The bosonic field equations in eleven dimensions are 
\cite{Cremmer:1978km}
\beq
R_{MN}-\thalf G_{MN} R=\tfrac{1}{3}(F_M{}^{PQR}F_{NPQR} - \tfrac{1}{8}G_{MN}F_{PQRS}F^{PQRS}),
\eeq
\beq
\nabla_MF^{MNPQ}=-\tfrac{1}{576}\epsilon^{NPQM_1...M_8}F_{M_1..M_4}F_{M_5..M_8},
\eeq
and the Bianchi identities read
\beq
\partial_{[M}F_{NPQR]}=0.
\eeq

Using a product  metric ansatz for the background splitting eleven dimensions into $4+7$ ($M=(\mu,m)$ etc), and a non-zero background value only for the spacetime components of the four-form  field strength, that is  \cite{Freund:1980xh}
\beq
\la G_{MN}\ra = 
 \begin{pmatrix}
  \bar g_{\mu\nu}&  0 \\
   0& \bar g_{mn}
  \end{pmatrix}
  ,\,\, \la F_{\mu\nu\rho\si}\ra =3m\bar{\epsilon}_{\mu\nu\rho\si},
\eeq
where $m$ is a positive parameter with dimension mass, we find that the Ricci tensors  in the external and internal directions become, respectively,
\beq
\la R_{\mu\nu}\ra=-12m^2\bar g_{\mu\nu},\,\, \la R_{mn}\ra =6m^2\bar g_{mn},
\eeq
where background values are indicated by a bar over the field in question.

The seven-sphere compactifications (round and squashed) of eleven-dimensional supergravity is a very well-studied subject, see, e.g., \cite{Duff:1986hr}.
The spectrum is obtained by 
linearising the field equations in eleven dimensions
followed by a diagonalisation of the coupled equations.  This leads to a number of relations between the mass operators ($M^2$ for bosonic fields in $AdS_4$ and $M$ for fermionic ones) and operators on the 
seven-sphere as given below in Table 2. From Table 1 one can then read off which irreps $D(E_0,s)$ of $SO(3,2)$ the eigenvalues of the mass operators correspond to up to the sign ambiguity for scalars and spin $1/2$ fermions\footnote{Imposing  supersymmetry may in some cases remove this ambiguity \cite{Hawking:1983mx}.}.

The spectra obtained in such compactifications are therefore given in terms of towers of irreps of the isometry groups $SO(3,2)\times SO(8)$ and $SO(3,2)\times (Sp(2)\times Sp(1))$ for the cases of interest here.
The $AdS_4$ fields transform in $SO(3,2)$ irreps $D(E_0,s)$ for spins $s=0, \thalf, 1, \tfrac{3}{2}, 2$ and some parity (see Table 2) with $E_0$ values constrained by unitarity as $E_0\geq s+\thalf$ for matter fields and $E_0\geq s+1$ for gauge fields. In addition scalar fields have masses restricted by the Breitenlohner-Freedman condition $M^2 \geq -m^2$ \cite{Breitenlohner:1982bm}. 
For scalars the minus sign in $E_0$ is therefore only relevant for masses in the range $3m^2 \geq M^2 \geq -m^2$. This is summarised in Table 1 \cite{Heidenreich:1982rz, Englert:1983rn, Sezgin:1983ik, Casher:1984ym}.
\begin{table}[H]
\begin{alignat}{2}
    E_0 &= \frac{3}{2} \pm \frac{1}{2}\sqrt{(M/m)^2 +1}\geq \thalf,& &\qquad{s=0,} \\
  E_0 &= \frac{3}{2} \pm \frac{1}{2}|{M/m}|\geq 1,& &\qquad{s=\frac{1}{2},} \\
 E_0 &= \frac{3}{2} + \frac{1}{2}\sqrt{(M/m)^2 +1 }\geq 2,& &\qquad{s=1,} \\
 E_0 &= \frac{3}{2} +\frac{1}{2}|{M/m-2}|\geq \tfrac{5}{2},& &\qquad{s=\frac{3}{2},} \\
  E_0 &= \frac{3}{2} + \frac{1}{2}\sqrt{(M/m)^2+9}\geq 3,& &\qquad{s=2.}
\end{alignat}
\caption{$E_0$ for $AdS_4$ fields of given mass $M$ and spin $s$ (in $SO(3,2)$ irreps $D(E_0,s)$) and the corresponding unitarity bounds.}
\end{table}

The singletons that are of interest in this paper are the two irreps $D(\thalf,0)$ and $D(1,\thalf)$ that saturate the unitary bounds for the two matter fields of spin zero and one half, respectively. Both of them rely
for their existence on the possibility to choose the minus sign in the above expressions for $E_0$. For the gauge fields the unitary bounds correspond to massless fields.

\vspace{5mm}
\begin{table}[H]
\vspace{5mm}
  \centering
   $ \begin{array}{l l} \toprule
    s & \text{Mass operator}  \\ \midrule
    2^{{+}} & \Delta_0 \\
    \tfrac{3}{2}^{\scriptscriptstyle (1),(2)} & \slashed{D}_{1/2} + \tfrac{7m}{2} \\
    1^{-\scriptscriptstyle (1),(2)} & \Delta_1 + 12m^2 \pm 6m\sqrt{\Delta_1 + 4m^2} \\
    1^+ & \Delta_2 \\
    \thalf^{\scriptscriptstyle (4),(1)} & \slashed{D}_{1/2} - \tfrac{9m}{2} \\
    \thalf^{\scriptscriptstyle (3),(2)} & \tfrac{3m}{2} - \slashed{D}_{3/2} \\
    0^{+ \scriptscriptstyle (1), (3)} & \Delta_0 + 44m^2 \pm 12m\sqrt{\Delta_0 + 9m^2} \\
    0^{+\scriptscriptstyle (2)} & \Delta_{L} - 4m^2 \\
    0^{-\scriptscriptstyle (1),(2)} & Q^2 + 6mQ + 8m^2 \\
    \bottomrule
    \end{array}$   
    \label{table:massop}
    \caption{Mass operators appearing in the Freund-Rubin  compactifications. The superscript signs on bosonic fields specify its parity. For spins $1^-$ and $0^+$ the minus sign in $M^2$ should be selected for the superscript $(1)$ and the plus sign for the second superscript. For spins $\tfrac{3}{2}, \thalf$ and $0^-$ the first superscript corresponds to the negative part of the spectrum of the linear operator in 
    question (i.e., $\slashed{D}_{1/2}, \slashed{D}_{3/2}$ and $Q$) while the second label corresponds to the positive part of the spectrum.}

\end{table}

The $AdS_4$  irreps were combined into ${\mathcal N}=1$ multiplets by Heidenreich  \cite{Heidenreich:1982rz} and into ${\mathcal N}=8$ multiplets by Freedman and Nicolai  \cite{Freedman:1983na} (see also the review by Nicolai \cite{Nicolai:1984hb}). These results are implicit in the discussions in the following subsections.

\subsection{The spectrum on the round  seven-sphere}

The spectrum on the round seven-sphere has been derived by several different methods, e.g., using the supergroup $OSp(4/N)$ or Young Tableaux, giving the result in Table 3.
By inserting the eigenvalue spectra of the operators in Table 3 into the mass operators of Table 2 one arrives at the ${\mathcal N}=8$ supermultiplet spectra presented in Table 4 where $n\geq 0$
refers to the level with $n=0$ giving  ${\mathcal N}=8$ supergravity  and $n\geq 1$  massive supermultiplets.

Since singletons play a key role in this paper we emphasise here that,
as discussed in \cite{Duff:1986hr}, spin-0 singletons arise in the $0^{(1)}$ tower
(see Table 2) when the scalar Laplacian $\Delta_0$ on the 7-manifold has
modes with eigenvalue $\tfrac{7m}{2}$.  These can occur only for the round $S^7$ (see,
for example, \cite{Duff:1984sv}), and so only the round $S^7$ vacuum can have spin-0
singletons.

In anticipation of the discussion of the squashed sphere spectrum  in the next subsection we note that 
spin-$\thalf$ singletons arise in the $\thalf^{(1)}$  tower (see Table 2) when the
Dirac operator on the 7-manifold has modes with eigenvalue $+\tfrac{7m}{2}$.  On
the round sphere there is an $8_c$ of such modes, together with an $8_s$ of
modes with eigenvalue $-\tfrac{7m}{2}$ that give the 8 supersymmetries of the round
vacuum.  With the exception of the round $S^7$, Dirac modes with eigenvalues
$+\tfrac{7m}{2}$ and $-\tfrac{7m}{2}$ cannot co-exist on any manifold (see \cite{Duff:1984sv}). So at most,
for any other manifold one can have either supersymmetry but no spin-$\thalf$
singletons or else spin-$\thalf$ singletons but no supersymmetry which is precisely what we will argue happens for the 
left and right squashed spheres, respectively.

\begin{table}[H]
    \centering
    $\begin{array}{l l l l}
         SO(7)\,\,\,\,& \text{Operator}\,\,\,\,& \text{SO(8) Dynkin label}\,\,\,\,\,\,& \text{Eigenvalues} \\
      {\bf 1}& \Delta_0& (p,0,0,0)& p(p+6)m^2 \\
       {\bf 7}& \Delta_1&  (p-1,1,0,0)& [p(p+6)+5]m^2 \\
        {\bf 21}& \Delta_2& (p-1,0,1,1)& [p(p+6)+8]m^2 \\
         {\bf 35}&  Q&       (p-1,0,2,0)& -(p+3)m \\
                        &   &       (p-1,0,0,2)& +(p+3)m \\
        {\bf 27}& \Delta_L& (p-2,2,0,0)&         [p(p+6)+12]m^2 \\   
        {\bf 8}& \slashed{D}_{1/2}& (p,0,1,0)&     +(p+\tfrac{7}{2})m \\
                       &                  & (p,0,0,1)&     -(p+\tfrac{7}{2})m \\
        {\bf 48}& \slashed{D}_{3/2}& (p-1,1,1,0)& +(p+\tfrac{7}{2})m \\
                        &                  & (p-1,1,0,1)& -(p+\tfrac{7}{2})m
    \end{array}$
    \caption{The round $S^7$ eigenvalues of the differential operators acting on $SO(8)$ harmonics. The $SO(7)$ representations are given in terms of their dimension.}
    \label{table:s7spec}
\end{table}

\begin{table}[H]
    \centering
    $\begin{array}{l l l l l}
\text{Spin} \qquad& \text{SO(8) rep}\qquad \qquad& E_0  \quad\qquad\qquad& (\text{Mass})^2 \qquad\qquad&\text{Operator}\\ \hline
2 & (n,0,0,0) & (n+6)/2 \,\,\,& (n+3)-9 &\Delta_0\\
\tfrac{3}{2}^{(1)} & (n,0,0,1) & (n+5)/2 & n^2 & \slashed{D}_{\tfrac{1}{2}}<0 \\
\tfrac{3}{2}^{(2)} & (n-1,0,1,0) & (n+7)/2 & (n+6)^2 & \slashed{D}_{\tfrac{1}{2}} >0\\
1^{-(1)} & (n,1,0,0) & (n+4)/2 & (n+1)^2-1& \Delta_1 \ge 12 \\
1^{-(2)} & (n-2,1,0,0) & (n+8)/2 & (n+5)^2-1&\Delta_1\ge 12\\
1^{+} & (n-1, 0, 1, 1) & (n+6)/2 & (n+3)^2-1& \Delta_2 \\
\tfrac{1}{2}^{(4)} & (n-2,0,0,1) & (n+9)/2 & (n+6)^2&\slashed{D}_{\tfrac{1}{2}}<0 \\
\tfrac{1}{2}^{(1)} & (n+1,0,1,0) & (n+3)/2 & n^2&\slashed{D}_{\tfrac{1}{2}}>0 \\
\tfrac{1}{2}^{(2)} & (n-1, 1, 1, 0) & (n+5)/2 & (n+2)^2&\slashed{D}_{\tfrac{3}{2}} >0 \\
\tfrac{1}{2}^{(3)} & (n-2, 1, 0, 1) & (n+7)/2 & (n+4)^2&\slashed{D}_{\tfrac{3}{2}}<0\\
0^{+(1)} & (n+2,0,0,0) & (n+2)/2 & (n-1)^2-1&\Delta_0 \\
0^{+(3)} & (n-2, 0,0,0) & (n+10)/2 & (n+7)^2-1&\Delta_0 \\
0^{+(2)} & (n-2, 2, 0, 0) & (n+6)/2 & (n+3)^2-1&\Delta_L \\
0^{-(1)} & (n,0,2,0) & (n+4)/2 & (n+1)^2-1&Q<0\\
0^{-(2)} & (n-2, 0, 0, 2) & (n+8)/2 & (n+5)^2-1&Q>0
\end{array}$
    \caption{The complete spectrum of particles on the round sphere compactification of eleven-dimensional supergravity. Each integer $n$ gives an entire supermultiplet of particles. For the linear operators the positive and negative part of the spectrum is associated with different towers as indicated by the inequality signs.}
    \label{table:complete_s7_spectrum}
\end{table}

\subsection{The spectrum of irreps on the squashed seven-sphere}

For the left-squashed sphere the structure of the various isometry towers must be compatible with the ${\mathcal N}=1$ supersymmetry present in $AdS_4$ in this case.
The unitary  ${\mathcal N}=1$ supermultiplets were constructed by Heidenreich \cite{Heidenreich:1982rz}:


\begin{table}[H]
    \centering
    \begin{align*}
    &\textbf{Type A: }\text{Wess-Zumino multiplets for $E_0 > 1/2$} \nonumber \\ 
    &D(E_0, 0) \oplus D(E_0 + 1/2, 1/2) \oplus D(E_0 + 1, 0) \\
    &\textbf{Type B: }\text{Massive higher spin multiplets for $E_0 > s+1$, $s \ge 1/2$}  \nonumber \\ 
    &D(E_0, s) \oplus D(E_0 + 1/2, s+1/2) \oplus D(E_0 + 1/2, s-1/2) \oplus D(E_0 + 1, s)\\
    &\textbf{Type C: }\text{Massless higher spin multiplets for $s \ge 1/2$} \nonumber \\
    &D(s+1, s) \oplus D(s+3/2, s+1/2) \\
    &\textbf{Type D: }\text{Dirac singleton } \nonumber\\
    &D(1/2, 0) \oplus D(1, 1/2)
    \end{align*}
    \caption{${\mathcal N}=1$ supermultiplets.}
    \label{tab:my_label}
\end{table}

Since the supersymmetry parameter (as well as the gravitino) is an isometry singlet all member fields in a supermultiplet must transform under the same isometry irrep.
That the isometry irreps of all towers fit exactly into such ${\mathcal N}=1$ supermultiplets   has been verified in full detail in this work.

The method employed here to get the full spectrum directly on the squashed sphere is based on an application of the rules of the game
spelt out by Salam and Strathdee in \cite{Salam:1981xd}.
Consider again each operator  in Table 2 and their harmonics (eigenfunctions) but now on the coset $Sp_2 \times Sp^C_1/Sp^{A}_1 \times Sp^{B+C}_1$.
Here we have used the splitting $Sp_2 \rightarrow Sp_1^A \times Sp_1^B$ in order to be able to define the diagonal subgroup $Sp_1^{B+C}$ in the denominator
subgroup of the coset.
We will refer to this method
as the Young Tableau Method (YTM) which is essentially just a realisation of the Fourier analysis for coset spaces $G/H$. It has, e.g., been used by D'Auria and Fr\'e in connection
with compactification on spaces like $M^{pqr}$, see \cite{DAuria:1983rmi} and references therein (see also \cite{Duff:1986hr}). 

Thus  we first split the tangent space $SO(7)$ irrep of the squashed $S^7$ tensor/spinor field in question into  $H=Sp^{A}_1 \times Sp^{B+C}_1$ irreps $(m,n)$ and then, 
for each such irrep, we tabulate all $G=Sp_2 \times Sp^C_1$ irreps that in their decompositions under $H$ contain the $H$ irrep $(m,n)$ we consider. Once this is done we collect all the $G$ irreps 
for the $S^7$ field and remove the irreps
that correspond to longitudinal states (and possible other similar states) so that the purely  transverse spectrum is arrived at at the end. We will illustrate the procedure in detail for the
 Lichnerowitz operator $\Delta_L$ below and give the result for all other operators in the Appendix. As it turns out, the spectrum of $\Delta_L$ together with that of $\slashed D_{3/2}$ contain all the 
 crucial features that will be used in the final argument for the need to incorporate singletons in the round sphere spectrum as well as in the one for $RS^7$. We will now explain the procedure  in a number of separate steps.
 
{\bf Step 1:} To start with we need to decompose of the various tangent space tensors into irreps of the coset denominator group $Sp_1^A\oplus Sp_1^{B+C}$. As it happens, by using the 
McKay and Patera tables \cite{McKayPatera},
 this is most easily done by a two-step decomposition via $G_2$ as follows: $SO(7)\rightarrow G_2 \rightarrow Sp_1^A\times Sp_1^{B+C}$. We find
 
\begin{align} 
\text{scalar}({\bf 1}) & :\,\, (000)\rightarrow  (00)\rightarrow (0,0),\\
\text{1-form}({\bf 7}) & :\,\,(100)\rightarrow  (01) \rightarrow (1,1)\oplus (0,2),\\
\text{Dirac}({\bf 8}) & :\,\,(001)\rightarrow  (01)\oplus (00)\rightarrow (1,1)\oplus (0,2)\oplus (0,0),\\
\text{2-form}({\bf 21}) & :\,\,(010)\rightarrow  (01)\oplus (10) \rightarrow (1,1)\oplus (0,2)\oplus (0,2) \oplus (1,3) \oplus (2,0),\\
\text{metric}({\bf 27}) & :\,\,(200)\rightarrow  (02)\rightarrow  (2,2)\oplus (1,3) \oplus (0,4) \oplus (1,1)\oplus (0,0),\\
\text{3-form}({\bf 35}) & :\,\,(002)\rightarrow  (02)\oplus (01) \oplus (00)\rightarrow  \text{see above},\\
\text{Rarita-Schw.}({\bf 48}) & :\,\,(101)\rightarrow  (02)\oplus (10) \oplus (01)\rightarrow \text{see above}.
\end{align}

Here the irreps are given in terms of Dynkin labels  except for the numbers in bold which refer to the  dimension of the irrep the corresponding field on $S^7$ belongs to.

 We now concentrate on the harmonics of the operator  $\Delta_L$, that is  metric harmonics. \\
 
 {\bf Step 2:}. The  traceless metric on the squashed seven-sphere is in the ${\bf 27}$ of  $SO(7)$ which splits as just mentioned into $H=Sp^{A}_1 \times Sp^{B+C}_1$ irreps $(m,n)$ (via $G_2$) as follows:
 \beq
 {\bf 27}: \,\,(200)\rightarrow  (02)\rightarrow  (2,2)\oplus (1,3) \oplus (0,4) \oplus (1,1)\oplus (0,0).
 \eeq
 
 {\bf Step 3:} For each of the $H$ irreps $(m,n)$ on the right hand side in the last equation we now look for all $G$ irreps that have the $H$ irrep in its decomposition.
 This is best done using Young Tableaux (YTs).  A general irrep of $G=Sp_2 \times Sp^C_1$ can be parametrised by three non-negative integers $(p,q;r)$ related to
 a  $Sp_2$ YT with $q$ columns with two boxes and $p$ columns of single boxes. The dimension of such $Sp_2$ irreps is given by $d(p,q)=\tfrac{1}{6}(p+1)(q+1)(p+q+2)(p+2q+3)$. 
 The one-box YT is thus denoted $(1,0)$ and has dimension four while one two-box column is $(0,1)$ with dimension five.
 Each $Sp_2$ irrep must then be combined with the irrep $(r)$ of the second group factor of $G$, i.e., $Sp_1^C$ having one row YTs with $r$ boxes. 
 A general  $(p,q;r)$ YT of this kind thus has the following form:
 
 \begin{equation}
\begin{ytableau}
{} & \none[ \dots ] & {} &  \tikzmark{startB} & \none[ \dots ] & \tikzmark{endB} \\
 \tikzmark{startA} & \none[ \dots ] & \tikzmark{endA} \\
 \none
\end{ytableau}
~\times~
\begin{ytableau}
\tikzmark{startC} & \none[ \dots ] & \tikzmark{endC}
\end{ytableau}
\begin{tikzpicture}[remember picture,overlay, baseline=(current  bounding  box.center)]
\draw[decorate,decoration={brace,raise=10pt, mirror}] 
  ([xshift=-8pt]{{pic cs:startA}|-{pic cs:endA}}) -- node[below=11pt] {$q$} ([xshift=2pt]pic cs:endA);
\draw[decorate,decoration={brace,raise=10pt, mirror}] 
  ([xshift=-5pt]{{pic cs:startB}|-{pic cs:endB}}) -- node[below=11pt] {$p$} ([xshift=3pt]pic cs:endB);
\draw[decorate,decoration={brace,raise=10pt, mirror}] 
  ([xshift=-5pt]{{pic cs:startC}|-{pic cs:endC}}) -- node[below=11pt] {$r$} ([xshift=3pt]pic cs:endC);
\end{tikzpicture}
\label{YT:sp2}
\end{equation}

 Consider for example\footnote{Note that in the list of $H$ irreps $(m,n)$ appearing in this context $m+n$ is always an even integer.} the $H$ irrep $(m,n)=(1,3)$ which we will think of as a tensor with one undotted and three symmetrised dotted  indices (all two-dimensional)\footnote{This stems from the decomposition of the index $A$ for the irrep ${\bf 4}$ of $Sp_2$ into $Sp_1\times Sp_1$ as $A=(a,\dot a)$.}  where the dotted indices arise from forming
 the diagonal subgroup of  $Sp_1^B$ and $Sp_1^C$. To see if this $H$ irrep is part of the decomposition of a particular $G$ irrep we first check if the $H$ irrep can be obtained by  filling the $Sp_2\times Sp_1^C$ YT with undotted and dotted indices using the fact that dotted indices represent  the diagonal subgroup $Sp_1^{B+C}$ irrep. Here we must keep in mind that  that each column filled with two indices of the same type is a singlet
 and hence does not contribute to the final $H$ irrep.
 
 As an illustration we choose the $G=Sp_2\times Sp_1^C$ irrep $(p,q;r)=(5,2;3)$. Then the single undotted index of $(1,3)$ can arise only in two ways, namely either by filling the single box columns with one undotted index and the remaining boxes with  dotted ones (and the same kind of index in all double-box columns) or by filling one of the boxes in one double-box column together with  all single box columns with dotted indices.  This gives the undotted content of this $H$ irrep. The dotted index content is then determined by forming the diagonal subgroup with the $(3)$ irrep of $Sp_1^C$ (which of course is also filled with dotted indices). This leads to the following two cases:\\

YT1:\,\,\young(\hfill\hfill\hfill\cdot\cdot\cdot\cdot,\hfill\hfill) x \young(\cdot \cdot\cdot)\, ,\hspace{10mm}
YT2:\,\,\young(\hfill\cdot \cdot\cdot\cdot\cdot\cdot,\hfill\hfill) x \young(\cdot \cdot\cdot).\\

\noindent where empty boxes are considered to be filled with undotted indices and those with a dot with dotted indices.

It is clear that the irrep $(3)$ of $Sp_1^{B+C}$ (i.e., three dotted indices) is present in the $Sp_1$ tensor product in both cases (since $(4) \otimes (3) = (1)\oplus (3) \oplus (5) \oplus (7)$ and similarly for the second case).

The crucial next step is to consider general $G$ irreps $(p,q;r)$ and decide which will contain the looked for $H$ irrep. In the example considered above, namely the $H$ irrep $(1,3)$,
we find that the first $Sp_2$ YT can be extended to any $q\geq 0$ while for the second YT we get the restriction $q\geq 1$. Finally we must determine the relation between the two
integers $p$ and $r$ so that their tensor product contains the irrep ${(3)}$. Clearly this gives the following eight cases (equal to the dimension of the $H$ irrep $(1,3)$):

YT1: $p\geq 1, q\geq 0$, and $r=p+2,\,r=p,\,r=p-2,\,$or$ \,\,r=p-4,$

YT2: $p\geq 0, q\geq 1$ and $r=p+4,\,r=p+2,\,r=p\,,\,\,$or$\,\,r=p-2.$

The final step to determine the possible values for the three integers $p,q,r$ is to check if the low integer cases really occur. This is not the case in general and we find that the eight cases
must be given individual lower bounds on $p$. The final result is\\
\beq
\noindent  YT1:
\eeq
 $q\geq 0$  with $(r=p+2, p\geq 1)$, $(r=p, p\geq 2)$, $(r=p-2, p\geq 3)$ or $(r=p-4, p\geq 4)$,\\
\beq
\noindent  YT2:
\eeq
 $q\geq 1$  with $(r=p+4, p\geq 0)$, $(r=p+2, p\geq 0)$, $(r=p, p\geq 1)$ or $(r=p-2, p\geq 2)$.\\

To facilitate the presentation of these eight infinite sets (towers)  of $G$ irreps for the $H$ irrep $(1,3)$, as well as all the other cases appearing in the above decomposition of the tangent space irreps,
we will use "tower diagrams". In the example discussed above the eight cases look like:\footnote{Note that the diagrams are supposed to be extended to infinity in the $p$ and $q$ 
directions as suggested by the crosses displayed.}\\

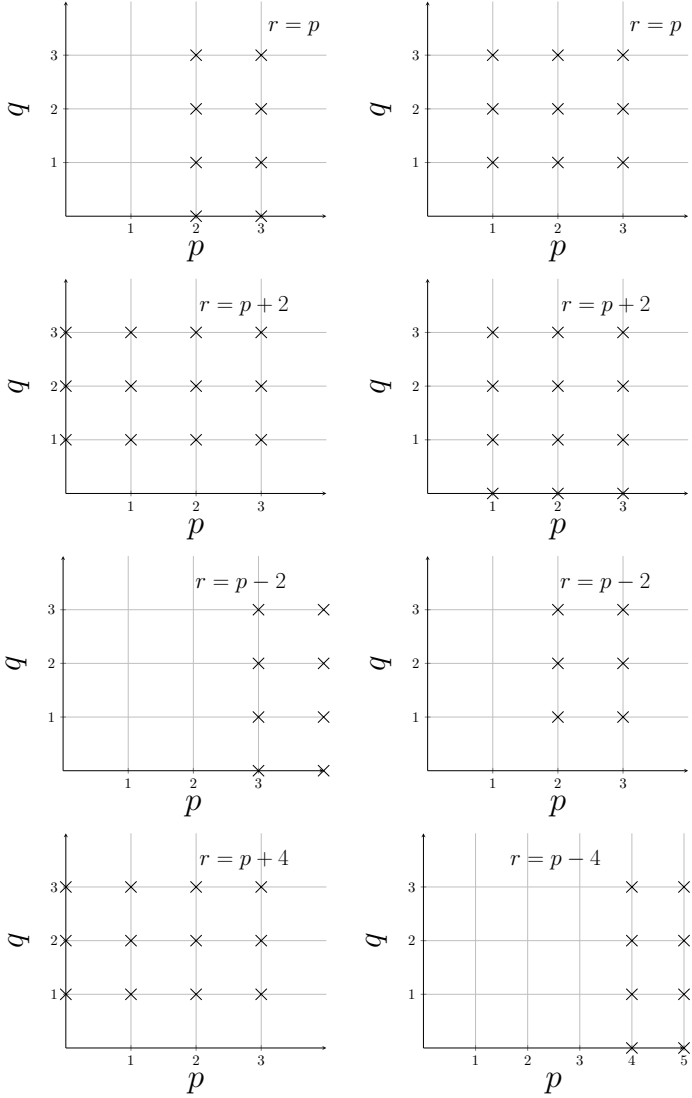
\begin{tabular}{c c c c}
\begin{tikzpicture}[scale=0.5]
\begin{axis}[%
    axis lines = left,
    xmin = 0, xmax = 4,
    ymin = 0, ymax = 4,
    xlabel = {\Huge $p$}, ylabel = {\Huge $q$},
    xtick={1,2,3}, ytick={1,2,3},
    xmajorgrids=true, ymajorgrids=true,
]
\addplot [only marks, mark=x, mark size=6pt] table {
2   0
2   1
2   2
3   0
3   1
3   2
2   3
3   3
};
\node at (axis cs:3.5,3.5) [anchor=center] {\LARGE $r=p$};
\end{axis}
\end{tikzpicture}
&
\begin{tikzpicture}[scale=0.5]
\begin{axis}[%
    axis lines = left,
    xmin = 0, xmax = 4,
    ymin = 0, ymax = 4,
    xlabel = {\Huge $p$}, ylabel = {\Huge $q$},
    xtick={1,2,3}, ytick={1,2,3},
    xmajorgrids=true, ymajorgrids=true,
]
\addplot [only marks, mark=x, mark size=6pt] table {
1   1
1   2
1   3
2   1
2   2
2   3
3   1
3   2
3   3
};
\node at (axis cs:3.5,3.5) [anchor=center] {\LARGE $r=p$};
\end{axis}
\end{tikzpicture}
\\
\begin{tikzpicture}[scale=0.5]
\begin{axis}[%
    axis lines = left,
    xmin = 0, xmax = 4,
    ymin = 0, ymax = 4,
    xlabel = {\Huge $p$}, ylabel = {\Huge $q$},
    xtick={1,2,3}, ytick={1,2,3},
    xmajorgrids=true, ymajorgrids=true,
]
\addplot [only marks, mark=x, mark size=6pt] table {
0   1
0   2
0   3
1   1
1   2
1   3
2   1
2   2
2   3
3   1
3   2
3   3
};
\node at (axis cs:3.5,3.5) [anchor=east] {\LARGE $r=p+2$};
\end{axis}
\end{tikzpicture}
&
\begin{tikzpicture}[scale=0.5]
\begin{axis}[%
    axis lines = left,
    xmin = 0, xmax = 4,
    ymin = 0, ymax = 4,
    xlabel = {\Huge $p$}, ylabel = {\Huge $q$},
    xtick={1,2,3}, ytick={1,2,3},
    xmajorgrids=true, ymajorgrids=true,
]
\addplot [only marks, mark=x, mark size=6pt] table {
1   0
1   1
1   2
1   3
2   0
2   1
2   2
2   3
3   0
3   1
3   2
3   3
};
\node at (axis cs:3.5,3.5) [anchor=east] {\LARGE $r=p+2$};
\end{axis}
\end{tikzpicture}
\\
\begin{tikzpicture}[scale=0.5]
\begin{axis}[%
    axis lines = left,
    xmin = 0, xmax = 4,
    ymin = 0, ymax = 4,
    xlabel = {\Huge $p$}, ylabel = {\Huge $q$},
    xtick={1,2,3}, ytick={1,2,3},
    xmajorgrids=true, ymajorgrids=true,
]
\addplot [only marks, mark=x, mark size=6pt] table {
3   0
3   1
3   2
3   3
4   0
4   1
4   2
4   3
};
\node at (axis cs:3.5,3.5) [anchor=east] {\LARGE $r=p-2$};
\end{axis}
\end{tikzpicture}
&
\begin{tikzpicture}[scale=0.5]
\begin{axis}[%
    axis lines = left,
    xmin = 0, xmax = 4,
    ymin = 0, ymax = 4,
    xlabel = {\Huge $p$}, ylabel = {\Huge $q$},
    xtick={1,2,3}, ytick={1,2,3},
    xmajorgrids=true, ymajorgrids=true,
]
\addplot [only marks, mark=x, mark size=6pt] table {
2   1
2   2
2   3
3   1
3   2
3   3
};
\node at (axis cs:3.5,3.5) [anchor=east] {\LARGE $r=p-2$};
\end{axis}
\end{tikzpicture}
\\
\begin{tikzpicture}[scale=0.5]
\begin{axis}[%
    axis lines = left,
    xmin = 0, xmax = 4,
    ymin = 0, ymax = 4,
    xlabel = {\Huge $p$}, ylabel = {\Huge $q$},
    xtick={1,2,3}, ytick={1,2,3},
    xmajorgrids=true, ymajorgrids=true,
]
\addplot [only marks, mark=x, mark size=6pt] table {
0   1
0   2
0   3
1   1
1   2
1   3
2   1
2   2
2   3
3   1
3   2
3   3
};
\node at (axis cs:3.5,3.5) [anchor=east] {\LARGE $r=p+4$};
\end{axis}
\end{tikzpicture}
&
\begin{tikzpicture}[scale=0.5]
\begin{axis}[%
    axis lines = left,
    xmin = 0, xmax = 5,
    ymin = 0, ymax = 4,
    xlabel = {\Huge $p$}, ylabel = {\Huge $q$},
    xtick={1,2,3,4,5}, ytick={1,2,3},
    xmajorgrids=true, ymajorgrids=true,
]
\addplot [only marks, mark=x, mark size=6pt] table {
4   0
4   1
4   2
4   3
5   0
5   1
5   2
5   3
};
\node at (axis cs:3.5,3.5) [anchor=east] {\LARGE $r=p-4$};
\end{axis}
\end{tikzpicture}
\end{tabular}%

We end this subsection by presenting the full supermultiplet content on $LS^7$, the squashed seven-sphere with ${\mathcal N}=1$ supersymmetry. 
Thus we put all the $AdS_4$ fields obtained in the compactification on the left-squashed seven-sphere
into ${\mathcal N}=1$ supermultiplets as given by Heidenreich \cite{Heidenreich:1982rz}. The information needed for this is provided in Appendix B of this paper.
Each supermultiplet is specified by the spin and parity of the field in $AdS_4$ with highest spin together with its isometry irrep (which is the same for all fields in the supermultiplet).
Note that we have here rearranged the Heidenreich supermultiplets  by ordering the irreps from largest to lowest spin $s$ as follows
\beq
D(E_0,s)\oplus D(E_0+\thalf, s-\thalf)\oplus D(E_0-\thalf, s-\thalf)\oplus D(E_0, s-1).
\eeq
For instance, the massive graviton supermultiplet in isometry irrep $(1,0;1)=({\bf 4},{\bf 2})$ then reads, with $E_0=\tthalf+\tothird\sqrt{11}$,
\beq
D(E_0,2^+)\oplus D(E_0+\thalf, \tfrac{3}{2})\oplus D(E_0-\thalf, \tthalf)\oplus D(E_0, 1^+).
\eeq

The total multiplet content of the $LS^7$ theory is (with massless multiplets being special cases in the relevant tower diagrams)
\beq
1\times (D(E_0,2^+)\oplus D(E_0+\thalf, \tfrac{3}{2})\oplus D(E_0-\thalf, \tthalf)\oplus D(E_0, 1^+)),
\eeq
\beq
6\times (D(E_0,\tthalf)\oplus D(E_0+\thalf, 1^{\pm})\oplus D(E_0-\thalf, 1^{\mp})\oplus D(E_0, \thalf)),
\eeq
\beq
6\times (D(E_0,1^-)\oplus D(E_0+\thalf, \thalf)\oplus D(E_0-\thalf, \thalf)\oplus D(E_0, 0^-)),
\eeq
\beq
8\times (D(E_0,1^+)\oplus D(E_0+\thalf, \thalf)\oplus D(E_0-\thalf, \thalf)\oplus D(E_0, 0^+)),
\eeq
\beq
14\times (D(E_0, \thalf)\oplus D(E_0+\thalf, 0^{\pm})\oplus D(E_0-\thalf, 0^{\mp})),
\eeq
where the multiplicity (number in front) refers to the number of tower diagrams that appear for the given supermultiplet. It should be emphasised that not only are all tower diagrams accounted for by the list above but in fact each individual irrep (i.e., cross) is given a place in a supermultiplet.

Note that in the last case, the Wess-Zumino multiplets, we have not committed ourselves to the parity assignment of the scalar fields. This can, however,  be done for the spin 1 fields in the six spin $3/2$
multiplets since we know exactly their $E_0$ values\footnote{There are three of each choice of signs.}.
Although we have full knowledge of the operator spectra on the squashed sphere for some operators ($\Delta_0, \Delta_1$ and $\slashed D_{1/2}$, see \cite{Duff:1986hr}) we lack
this for the other operators (although partial results exist) which means that some supermultiplets among the last two categories in the above list cannot unambiguously be assigned
values of $E_0$. 

The known massless supermultiplets are the graviton supermultiplet
\beq
\text{Spin}\,\, 2^+(0,0;0)=({\bf 1},{\bf 1}):\,\,D(3,2)\oplus D(\tfrac{5}{2},\tfrac{3}{2})
\eeq
and the two gauge supermultiplets
\beq
\text{Spin} \,\,1^-(0,0;2)=({\bf 1},{\bf 3}):\,\,D(2,1)\oplus D(\tfrac{3}{2},\tfrac{1}{2})
\eeq
\beq
\text{Spin} \,\,1^-(2,0;0)=({\bf 10},{\bf 1}):\,\,D(2,1)\oplus D(\tfrac{3}{2},\tfrac{1}{2})
\eeq
The only other kind of massless supermultiplets that can appear are Wess-Zumino ones. However, we need more detailed information about the more complicated operators
($\Delta_L$,  $Q$ and ${\slashed D}_{3/2}$) to determine whether or not they  occur.

\subsection{Comparison of the irrep spectra  on the round and squashed seven-spheres}

The supergravity theory obtained by compactifying on the left-squashed seven-sphere has one (four-dimensional) supersymmetry,
i.e.,  ${\mathcal N} =1$, and is believed to arise as a spontaneously broken version of the round sphere supergravity theory with ${\mathcal N} =8$.
In this spontaneous symmetry breaking (SSB) the isometry is broken as $SO(8)\rightarrow Sp_2\times Sp_1$ with two possible results depending how
 the broken isometry group is embedded into the unbroken one. The two options are defined by stating how the three eight-dimensional irreps
of $SO(8)$ break: For the left-squashed case $LS^7$
\beq
S^7 \rightarrow LS^7: {\bf 8}_v\rightarrow ({\bf 4},{\bf 2}),\,\,{\bf 8}_s\rightarrow ({\bf 4},{\bf 2}),\,\,{\bf 8}_c\rightarrow ({\bf 5},{\bf 1})\oplus ({\bf 1},{\bf 3}),
\eeq
and for the right-squashed case $RS^7$
\beq
S^7 \rightarrow  RS^7: {\bf 8}_v\rightarrow ({\bf 4},{\bf 2}),\,\,{\bf 8}_s\rightarrow ({\bf 5},{\bf 1})\oplus ({\bf 1},{\bf 3}),\,\,{\bf 8}_c\rightarrow ({\bf 4},{\bf 2}) .
\eeq
In \cite{Duff:1986hr} this is discussed in some detail in particular in connection with the so called {\it space  invader scenario}. This refers to the fact that
the eight massless gravitini on the round sphere are replaced by a single gravitino on the left-squashed sphere despite the fact that
the symmetry breaking $SO(8)\rightarrow Sp(2)\times Sp(1)$ tells us that all the eight gravitini on the round sphere become massive after the breaking.
This follows immediately if we recall that they belong to the $SO(8)$ irrep ${\bf 8}_s$ on the round sphere and hence belong  to the irrep $(${\bf 4},{\bf 2}$)$ after the SSB.
This seemingly strange fact is explained by noting that there is after the breaking a singlet mode coming from the massive part of the round sphere spectrum which zooms down
and becomes massless in the left-squashed sphere spectrum.

Thus the space invader scenario involves two Higgs phenomena, one ordinary one in which the eight round sphere gravitini become massive by absorbing
a set of spin $\thalf$  fields in the same isometry representation which is $(${\bf 4},{\bf 2}$)$ after the spontaneous symmetry breaking (SSB).  The second kind of Higgsing is perhaps more appropriately 
called deHiggsing since  in this case a singlet massive  gravitino becomes massless by spitting out a spin $\thalf$ fermion field. In this paper we have 
performed a complete analysis of the irrep content of the spectrum on the $LS^7$ using a method that is completely independent of the round sphere theory and its connection to it via the SSB
described above. This is quite interesting in its own right but it is by trying to connect it to the SSB of the round sphere that certain special features are discovered.
As we will see below some of these features seem to tell us that the singleton representations that normally are gauged away and thus discarded on the round sphere (at least in the bulk)
must be kept as part of the round $S^7$ theory although they are strictly speaking confined to the boundary of $AdS_4$.

The $G$ irreps that do not match  in the comparison between the round and squashed  irrep spectra are

\beq
\text{for} \,\,\,\Delta_L:\,\,\, ({\bf 4},{\bf 2}), ({\bf 5},{\bf 3}),
\eeq
\beq
\text{for}\,\,\, {\slashed D}_{3/2}:\,\,\, ({\bf 1},{\bf 1})^*, ({\bf 4},{\bf 2}), ({\bf 5},{\bf 1}), ({\bf 1},{\bf 3}).
\eeq

Here we emphasise that all of these irreps except $({\bf 1},{\bf 1})$ (thus the *) arise from the $SSB$ of the round spectrum but do not occur in the squashed spectrum when it is derived directly by our YT methods as explained above. The natural way to understand this is through a Higgs mechanism as done in \cite{Duff:1986hr}. However, there this was only applied to the $\Delta_L$ mode $(${\bf 5},{\bf 3}$)$
and the ${\slashed D}_{3/2}$ mode  $(${\bf 4},{\bf 2}$)$\footnote{This was done  without knowing to which operators they belong on the squashed sphere.}: The $(${\bf 5},{\bf 3}$)$ bosonic scalar modes are eaten so that the gauge fields 
in this irrep can become massive on the squashed sphere, for both left and right squashing, while the fermionic $(${\bf 4},{\bf 2}$)$ is eaten by the Rarita-Schwinger fields so that they can become massive, again for both left and right squashing.
The bosonic $(${\bf 4},{\bf 2}$)$ and the  fermionic modes $(${\bf 5},{\bf 1}$), ({\bf 1},{\bf 3})$, on the other hand, do not get any obvious SSB explanation in \cite{Duff:1986hr} since they were, in fact, not identified as modes in this category.

This brings us to the novel aspect of this paper namely the proposition that the remaining modes should be viewed as some kind of Goldstone modes for the singletons that we now
incorporate into the round sphere spectrum. Since they break to $(${\bf 4},{\bf 2}$)$ for the spin zero singleton and to $(${\bf 5},{\bf 1}$)\oplus  (${\bf 1},{\bf 3}$)$ for the spin half one in the left-squashed case, their presence  explains why the remaining modes do not 
appear in the squashed spectrum namely because they are eaten by these singleton fields which then  become ordinary bulk fields. We should emphasise that these bulk fields do appear in the left-squashed spectrum.

Having provided a plausible $SSB$ explanation for the left-squashed excess modes (encircled in the $\Delta_L$ and $\slashed D_{3/2}$ tower diagrams in Appendix B) we now turn to the
${\slashed D}_{3/2}$  singlet mode $(${\bf 1},{\bf 1}$)$ which is lacking in the $SSB$ spectrum although it does appear in the direct YT construction of the squashed spectrum (indicated by a box with a cross in the ${\slashed D}_{3/2}$ tower diagram in Appendix B). On the left-squashed sphere this is naturally explained by a reversed Higgsing (or deHiggsing) of the singlet (after SSB) massive Rarita-Schwinger field on the round sphere that becomes massless
in the left-squashed case and hence must relieve itself of its spin $\thalf$ states which then appear by themselves in the squashed spectrum as we have found in this paper.

This gives a complete picture for the $SSB$  relation between the round and left-squashed spectra. Turning to the corresponding situation in the right-squashed case we note
that the only difference is the fact that the deHiggsing of the massive Rarita-Schwinger field on $LS^7$  does not take place on $RS^7$ since in this latter case there are no supersymmetries.
However, there is still the singlet fermionic mode in the  $SSB$ of the round ${\slashed D}_{3/2}$ spectrum that also  appears in the direct $YT$ construction of the spectrum.

In the spirit of this paper we therefore propose that also the right-squashed supergravity theory on $AdS_4$ must contain a singleton, but this time only a fermionic one in the irrep $(${\bf 1},{\bf 1}$)$, which, 
together with the mode we are looking for, is the result of a deHiggsing of an ordinary  bulk field.
This is also supported by the fact that this singleton is indeed present (see the mass operators in Table 2) since it corresponds to the ${\slashed D}_{1/2}$ Killing spinor mode that gives rise to the massless Rarita-Schwinger field on the left-squashed sphere. On the left-squashed sphere this mode satisfies ${\slashed D}_{1/2}=-\tfrac{7m}{2}$ which after skew-whiffing to the right-squashed case 
flips sign\footnote{All operators linear in derivatives flip sign under skew-whiffing which just corresponds to a reversal of the orientation of the squashed seven-sphere \cite{Duff:1983ajq, Duff:1986hr}.} to ${\slashed D}_{1/2}=+\tfrac{7m}{2}$ and hence gives rise to a $M=-m$ spin $\thalf$ fermionic field in $AdS_4$ which is a singleton irrep of $SO(3,2)$ if the minus sign is chosen for $s=\thalf$ in 
Table 2.  It is perhaps interesting to note in this context that the origin on the round sphere of this singlet fermionic mode is ${\bf 56}_s$ (which in the $RS^7$ context breaks as ${\bf 56}_c$ in $LS^7$)
which happens to have zero mass. Hence this mode has $E_0=\tfrac{3}{2}$ and the sign in Table 2 is irrelevant. Once supersymmetry is lost in the SSB to the right-squashed case the properties of the  
fermion\footnote{The properties relevant  here are discussed in detail in \cite{Breitenlohner:1982jf}.} becomes independent of the other fields
and may pick the minus sign in the formula for $E_0$. This way we have described a possible scenario based on SSB and Higgsing/deHiggsing that explains all the 
modes listed above\footnote{See \cite{Duff:2018xx} for more details on the singlet sector of the round and squashed sphere spectra.}.

\section{Conclusions}
The role of singletons in the context of seven-sphere compactification of eleven-dimensional supergravity
has been an intriguing subject for a long time. In this paper we have found indications that the ${\mathcal N}=8$ supersingleton
must be kept in the $AdS_4$ spectrum arising from compactification on the round $S^7$ in order that  its spectrum can be Higgsed/deHiggsed 
 to produce the spectrum
obtained from compactification on the ${\mathcal N}=1$ left-squashed sphere $LS^7$. 

The assumption about a Higgs relation between these two compactifications seems to explain all modes present in the two cases including their 
precise relation. The new ingredient that is needed for this picture to work is, however, a novel kind of Higgsing of singleton that turn a singleton
which in some sense lives on the boundary of $AdS_4$ into an ordinary bulk field (of the same spin) by "eating" another bulk field of the same spin.
As explained in Section 2, by adopting this point of view a mismatch between the spectra of the left-squashed and right-squashed $RS^7$ (with no supersymmetry)
can also be corrected giving further support for this singleton Higgsing picture. As it turns out, for this to work a fermionic singleton is required in the spectrum of $RS^7$ whose
existence can, in fact, be directly verified.

Although a singleton Higgs effect may be novel, one could interpret the different state diagrams for $SO(3,2)$ irreps $D(E_0,s)$ in, e.g., the Appendix of the review \cite{Nicolai:1984hb} as  indicating a singleton Higgs effect similar to the one hinted at in that review for spin 1 gauge fields.  If this can be made explicit in a field theory for singletons it would be quite interesting
but as far as we are aware nothing in this direction has been attempted so far.

If the scenario presented here is correct it might have implications for how we view for instance the connection between $AdS$ bulk theories and their $CFT$
duals as stated in the $AdS/CFT$ correspondence \cite{Maldacena:1997re}. This in particular could mean that singletons have two different roles to play in the
$AdS/CFT$ context, being present  both as the $CFT$ and as part of the $AdS$ bulk theory.

We may also mention that the incorporation of a fermionic singleton in the $RS^7$ spectrum does not affect its established stability properties. Despite the fact that it has no supersymmetry, it is  Breitenlohner-Freedman stable since it can be obtained as a skew-whiffed version of the $LS^7$ theory, see, e.g., \cite{Duff:1986hr}. Furthermore,
the question if there is a an instability due to marginal  bound states operators in the $CFT$ of the $RS^7$ theory (corresponding to tadpoles in the bulk) \cite{Berkooz:1998qp} is not affected either by the fermionic singleton (which has conformal dimension equal to one).

\section*{Acknowledgement}
We thank M.J. Duff for a number of discussions on issues related to 
the theories analysed in this paper. C.N.P. is partially supported by
DOE grant DE-FG02-13ER42020.


\appendix
\section{H-tower diagrams}

\begin{figure}[H]
    \begin{subfigure}[t]{0.4\linewidth}
    \centering
    \begin{tikzpicture}
\begin{axis}[
    axis lines = left,
    xmin = 0, xmax = 4,
    ymin = 0, ymax = 4,
    xlabel = {\Huge $p$}, ylabel = {\Huge $q$},
    xtick={1,2,3}, ytick={1,2,3},
    xmajorgrids=true, ymajorgrids=true,
]
\addplot [only marks, mark=x, mark size=6pt] table {
0   0
0   1
0   2
0   3
1   0
1   1
1   2
1   3
2   0
2   1
2   2
2   3
3   0
3   1
3   2
3   3
};
\node at (axis cs:3.5,3.5) [anchor=center] {\LARGE $r=p$};
\end{axis}
\end{tikzpicture}
    \caption{(0,0) tower}
    \label{fig:my_label}
    \end{subfigure}%
    \begin{subfigure}[t]{0.6\linewidth}
    \centering
    \begin{tabular}{c c c c}
\begin{tikzpicture}
\begin{axis}[%
    axis lines = left,
    xmin = 0, xmax = 4,
    ymin = 0, ymax = 4,
    xlabel = {\Huge $p$}, ylabel = {\Huge $q$},
    xtick={1,2,3}, ytick={1,2,3},
    xmajorgrids=true, ymajorgrids=true,
]
\addplot [only marks, mark=x, mark size=6pt] table {
1   0
1   1
1   2
1   3
2   0
2   1
2   2
2   3
3   0
3   1
3   2
3   3
};
\node at (axis cs:3.5,3.5) [anchor=center] {\LARGE $r=p$};
\end{axis}
\end{tikzpicture}
&
\begin{tikzpicture}
\begin{axis}[%
    axis lines = left,
    xmin = 0, xmax = 4,
    ymin = 0, ymax = 4,
    xlabel = {\Huge $p$}, ylabel = {\Huge $q$},
    xtick={1,2,3}, ytick={1,2,3},
    xmajorgrids=true, ymajorgrids=true,
]
\addplot [only marks, mark=x, mark size=6pt] table {
0   1
0   2
0   3
1   1
1   2
1   3
2   1
2   2
2   3
3   1
3   2
3   3
};
\node at (axis cs:3.5,3.5) [anchor=center] {\LARGE $r=p$};
\end{axis}
\end{tikzpicture}
\\
\begin{tikzpicture}
\begin{axis}[%
    axis lines = left,
    xmin = 0, xmax = 4,
    ymin = 0, ymax = 4,
    xlabel = {\Huge $p$}, ylabel = {\Huge $q$},
    xtick={1,2,3}, ytick={1,2,3},
    xmajorgrids=true, ymajorgrids=true,
]
\addplot [only marks, mark=x, mark size=6pt] table {
0   1
0   2
0   3
1   1
1   2
1   3
2   1
2   2
2   3
3   1
3   2
3   3
};
\node at (axis cs:3.5,3.5) [anchor=east] {\LARGE $r=p+2$};
\end{axis}
\end{tikzpicture}
&
\begin{tikzpicture}
\begin{axis}[%
    axis lines = left,
    xmin = 0, xmax = 4,
    ymin = 0, ymax = 4,
    xlabel = {\Huge $p$}, ylabel = {\Huge $q$},
    xtick={1,2,3}, ytick={1,2,3},
    xmajorgrids=true, ymajorgrids=true,
]
\addplot [only marks, mark=x, mark size=6pt] table {
2   0
2   1
2   2
2   3
3   0
3   1
3   2
3   3
};
\node at (axis cs:3.5,3.5) [anchor=east] {\LARGE $r=p-2$};
\end{axis}
\end{tikzpicture}
\end{tabular}%
    \caption{(1,1) towers}
    \label{fig:my_label}
    \end{subfigure}
    \caption{}
\end{figure}
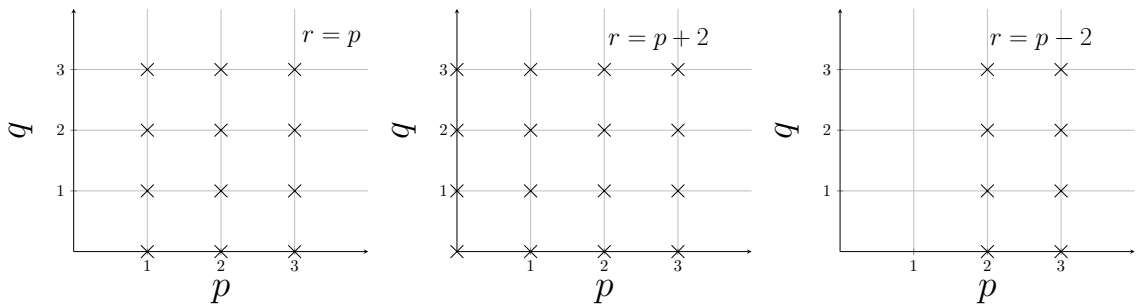
\begin{figure}[H]
    \centering
    \begin{tabular}{c c c c}
\begin{tikzpicture}
\begin{axis}[%
    axis lines = left,
    xmin = 0, xmax = 4,
    ymin = 0, ymax = 4,
    xlabel = {\Huge $p$}, ylabel = {\Huge $q$},
    xtick={1,2,3}, ytick={1,2,3},
    xmajorgrids=true, ymajorgrids=true,
]
\addplot [only marks, mark=x, mark size=6pt] table {
1   0
1   1
1   2
1   3
2   0
2   1
2   2
2   3
3   0
3   1
3   2
3   3
};
\node at (axis cs:3.5,3.5) [anchor=center] {\LARGE $r=p$};
\end{axis}
\end{tikzpicture}
&
\begin{tikzpicture}
\begin{axis}[%
    axis lines = left,
    xmin = 0, xmax = 4,
    ymin = 0, ymax = 4,
    xlabel = {\Huge $p$}, ylabel = {\Huge $q$},
    xtick={1,2,3}, ytick={1,2,3},
    xmajorgrids=true, ymajorgrids=true,
]
\addplot [only marks, mark=x, mark size=6pt] table {
0   0
0   1
0   2
0   3
1   0
1   1
1   2
1   3
2   0
2   1
2   2
2   3
3   0
3   1
3   2
3   3
};
\node at (axis cs:3.5,3.5) [anchor=east] {\LARGE $r=p+2$};
\end{axis}
\end{tikzpicture}
&
\begin{tikzpicture}
\begin{axis}[%
    axis lines = left,
    xmin = 0, xmax = 4,
    ymin = 0, ymax = 4,
    xlabel = {\Huge $p$}, ylabel = {\Huge $q$},
    xtick={1,2,3}, ytick={1,2,3},
    xmajorgrids=true, ymajorgrids=true,
]
\addplot [only marks, mark=x, mark size=6pt] table {
2   0
2   1
2   2
2   3
3   0
3   1
3   2
3   3
};
\node at (axis cs:3.5,3.5) [anchor=east] {\LARGE $r=p-2$};
\end{axis}%
\end{tikzpicture}%
\end{tabular}%
    \caption{(0,2) towers}
    \label{fig:my_label}
\end{figure}
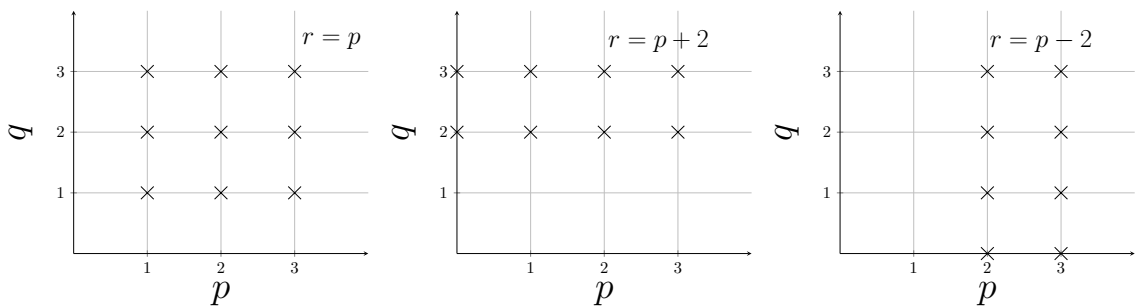
\begin{figure}[H]
    \centering
    \begin{tabular}{c c c c}
\begin{tikzpicture}
\begin{axis}[%
    axis lines = left,
    xmin = 0, xmax = 4,
    ymin = 0, ymax = 4,
    xlabel = {\Huge $p$}, ylabel = {\Huge $q$},
    xtick={1,2,3}, ytick={1,2,3},
    xmajorgrids=true, ymajorgrids=true,
]
\addplot [only marks, mark=x, mark size=6pt] table {
1   1
1   2
1   3
2   1
2   2
2   3
3   1
3   2
3   3
};
\node at (axis cs:3.5,3.5) [anchor=center] {\LARGE $r=p$};
\end{axis}
\end{tikzpicture}
&
\begin{tikzpicture}
\begin{axis}[%
    axis lines = left,
    xmin = 0, xmax = 4,
    ymin = 0, ymax = 4,
    xlabel = {\Huge $p$}, ylabel = {\Huge $q$},
    xtick={1,2,3}, ytick={1,2,3},
    xmajorgrids=true, ymajorgrids=true,
]
\addplot [only marks, mark=x, mark size=6pt] table {
0   2
0   3
1   2  
1   3
2   2
2   3
3   2
3   3
};
\node at (axis cs:3.5,3.5) [anchor=east] {\LARGE $r=p+2$};
\end{axis}
\end{tikzpicture}
&
\begin{tikzpicture}
\begin{axis}[%
    axis lines = left,
    xmin = 0, xmax = 4,
    ymin = 0, ymax = 4,
    xlabel = {\Huge $p$}, ylabel = {\Huge $q$},
    xtick={1,2,3}, ytick={1,2,3},
    xmajorgrids=true, ymajorgrids=true,
]
\addplot [only marks, mark=x, mark size=6pt] table {
2   0
2   1
2   2
2   3
3   0
3   1
3   2
3   3
};
\node at (axis cs:3.5,3.5) [anchor=east] {\LARGE $r=p-2$};
\end{axis}
\end{tikzpicture}
\end{tabular}%
    \caption{(2,0) towers}
    \label{fig:my_label}
\end{figure}
\begin{figure}    [H]
    \begin{subfigure}[t]{.5\linewidth}
    \centering
    \begin{tabular}{c c c c}
\begin{tikzpicture}
\begin{axis}[%
    axis lines = left,
    xmin = 0, xmax = 4,
    ymin = 0, ymax = 4,
    xlabel = {\Huge $p$}, ylabel = {\Huge $q$},
    xtick={1,2,3}, ytick={1,2,3},
    xmajorgrids=true, ymajorgrids=true,
]
\addplot [only marks, mark=x, mark size=6pt] table {
2   0
2   1
2   2
2   3
3   0
3   1
3   2
3   3
};
\node at (axis cs:3.5,3.5) [anchor=center] {\LARGE $r=p$};
\end{axis}
\end{tikzpicture}
&
\begin{tikzpicture}
\begin{axis}[%
    axis lines = left,
    xmin = 0, xmax = 4,
    ymin = 0, ymax = 4,
    xlabel = {\Huge $p$}, ylabel = {\Huge $q$},
    xtick={1,2,3}, ytick={1,2,3},
    xmajorgrids=true, ymajorgrids=true,
]
\addplot [only marks, mark=x, mark size=6pt] table {
1   1
1   2
1   3
2   1
2   2
2   3
3   1
3   2
3   3
};
\node at (axis cs:3.5,3.5) [anchor=center] {\LARGE $r=p$};
\end{axis}
\end{tikzpicture}
\\
\begin{tikzpicture}
\begin{axis}[%
    axis lines = left,
    xmin = 0, xmax = 4,
    ymin = 0, ymax = 4,
    xlabel = {\Huge $p$}, ylabel = {\Huge $q$},
    xtick={1,2,3}, ytick={1,2,3},
    xmajorgrids=true, ymajorgrids=true,
]
\addplot [only marks, mark=x, mark size=6pt] table {
0   1
0   2
0   3
1   3
1   1
1   2
2   1
2   2
2   3 
3   1
3   2
3   3
};
\node at (axis cs:3.5,3.5) [anchor=east] {\LARGE $r=p+2$};
\end{axis}
\end{tikzpicture}
&
\begin{tikzpicture}
\begin{axis}[%
    axis lines = left,
    xmin = 0, xmax = 4,
    ymin = 0, ymax = 4,
    xlabel = {\Huge $p$}, ylabel = {\Huge $q$},
    xtick={1,2,3}, ytick={1,2,3},
    xmajorgrids=true, ymajorgrids=true,
]
\addplot [only marks, mark=x, mark size=6pt] table {
1   0
1   1
1   2
1   3
2   0
2   1
2   2
2   3
3   0
3   1
3   2
3   3
};
\node at (axis cs:3.5,3.5) [anchor=east] {\LARGE $r=p+2$};
\end{axis}
\end{tikzpicture}
\\
\begin{tikzpicture}
\begin{axis}[%
    axis lines = left,
    xmin = 0, xmax = 4,
    ymin = 0, ymax = 4,
    xlabel = {\Huge $p$}, ylabel = {\Huge $q$},
    xtick={1,2,3, 4}, ytick={1,2,3},
    xmajorgrids=true, ymajorgrids=true,
]
\addplot [only marks, mark=x, mark size=6pt] table {
3   0
3   1
3   2
3   3
4   0
4   1
4   2
4   3
};
\node at (axis cs:3.5,3.5) [anchor=east] {\LARGE $r=p-2$};
\end{axis}
\end{tikzpicture}
&
\begin{tikzpicture}
\begin{axis}[%
    axis lines = left,
    xmin = 0, xmax = 4,
    ymin = 0, ymax = 4,
    xlabel = {\Huge $p$}, ylabel = {\Huge $q$},
    xtick={1,2,3}, ytick={1,2,3},
    xmajorgrids=true, ymajorgrids=true,
]
\addplot [only marks, mark=x, mark size=6pt] table {
2   1
2   2
2   3
3   1
3   2
3   3
};
\node at (axis cs:3.5,3.5) [anchor=east] {\LARGE $r=p-2$};
\end{axis}
\end{tikzpicture}
\\
\begin{tikzpicture}
\begin{axis}[%
    axis lines = left,
    xmin = 0, xmax = 4,
    ymin = 0, ymax = 4,
    xlabel = {\Huge $p$}, ylabel = {\Huge $q$},
    xtick={1,2,3}, ytick={1,2,3},
    xmajorgrids=true, ymajorgrids=true,
]
\addplot [only marks, mark=x, mark size=6pt] table {
0   1
0   2
0   3
1   1
1   2
1   3
2   1
2   2
2   3
3   1
3   2
3   2
3   3
};
\node at (axis cs:3.5,3.5) [anchor=east] {\LARGE $r=p+4$};
\end{axis}
\end{tikzpicture}
&
\begin{tikzpicture}
\begin{axis}[%
    axis lines = left,
    xmin = 0, xmax = 5,
    ymin = 0, ymax = 4,
    xlabel = {\Huge $p$}, ylabel = {\Huge $q$},
    xtick={1,2,3,4,5}, ytick={1,2,3},
    xmajorgrids=true, ymajorgrids=true,
]
\addplot [only marks, mark=x, mark size=6pt] table {
4   0
4   1
4   2
4   3
5   0
5   1
5   2
5   3
};
\node at (axis cs:3.5,3.5) [anchor=east] {\LARGE $r=p-4$};
\end{axis}
\end{tikzpicture}
\end{tabular}%
    \caption{(1,3) towers}
    \label{fig:my_label}
    \end{subfigure}
    \begin{subfigure}[t]{.5\linewidth}
    \centering
    \begin{tabular}{c c c c}
\begin{tikzpicture}
\begin{axis}[%
    axis lines = left,
    xmin = 0, xmax = 4,
    ymin = 0, ymax = 4,
    xlabel = {\Huge $p$}, ylabel = {\Huge $q$},
    xtick={1,2,3}, ytick={1,2,3},
    xmajorgrids=true, ymajorgrids=true,
]
\addplot [only marks, mark=x, mark size=6pt] table {
2   0
2   1
2   2
2   3
3   0
3   1
3   2
3   3
};
\node at (axis cs:3.5,3.5) [anchor=center] {\LARGE $r=p$};
\end{axis}
\end{tikzpicture}
&
\begin{tikzpicture}
\begin{axis}[%
    axis lines = left,
    xmin = 0, xmax = 4,
    ymin = 0, ymax = 4,
    xlabel = {\Huge $p$}, ylabel = {\Huge $q$},
    xtick={1,2,3}, ytick={1,2,3},
    xmajorgrids=true, ymajorgrids=true,
]
\addplot [only marks, mark=x, mark size=6pt] table {
1   0
1   1
1   2
1   3
2   0
2   1
2   2
2   3
3   0
3   1
3   2
3   3
};
\node at (axis cs:3.5,3.5) [anchor=east] {\LARGE $r=p+2$};
\end{axis}
\end{tikzpicture}
\\
\begin{tikzpicture}
\begin{axis}[%
    axis lines = left,
    xmin = 0, xmax = 4,
    ymin = 0, ymax = 4,
    xlabel = {\Huge $p$}, ylabel = {\Huge $q$},
    xtick={1,2,3,4}, ytick={1,2,3},
    xmajorgrids=true, ymajorgrids=true,
]
\addplot [only marks, mark=x, mark size=6pt] table {
3   0
3   1
3   2
3   3
4   0
4   1
4   2
4   3
};
\node at (axis cs:3.5,3.5) [anchor=east] {\LARGE $r=p-2$};
\end{axis}
\end{tikzpicture}
&
\begin{tikzpicture}
\begin{axis}[%
    axis lines = left,
    xmin = 0, xmax = 4,
    ymin = 0, ymax = 4,
    xlabel = {\Huge $p$}, ylabel = {\Huge $q$},
    xtick={1,2,3}, ytick={1,2,3},
    xmajorgrids=true, ymajorgrids=true,
]
\addplot [only marks, mark=x, mark size=6pt] table {
0   0
0   1
0   2
0   3
1   0
1   1
1   2
1   3
2   0
2   1
2   2
2   3
3   0
3   1
3   2
3   3
};
\node at (axis cs:3.5,3.5) [anchor=east] {\LARGE $r=p+4$};
\end{axis}
\end{tikzpicture}
\\
\begin{tikzpicture}
\begin{axis}[%
    axis lines = left,
    xmin = 0, xmax = 5,
    ymin = 0, ymax = 4,
    xlabel = {\Huge $p$}, ylabel = {\Huge $q$},
    xtick={1,2,3,4,5}, ytick={1,2,3},
    xmajorgrids=true, ymajorgrids=true,
]
\addplot [only marks, mark=x, mark size=6pt] table {
4   0
4   1
4   2
4   3
5   0
5   1
5   2
5   3
};
\node at (axis cs:3.5,3.5) [anchor=east] {\LARGE $r=p-4$};
\end{axis}
\end{tikzpicture}
\end{tabular}%
    \caption{(0,4) towers}
    \label{fig:my_label}
    \end{subfigure}
    \caption{}
\end{figure}
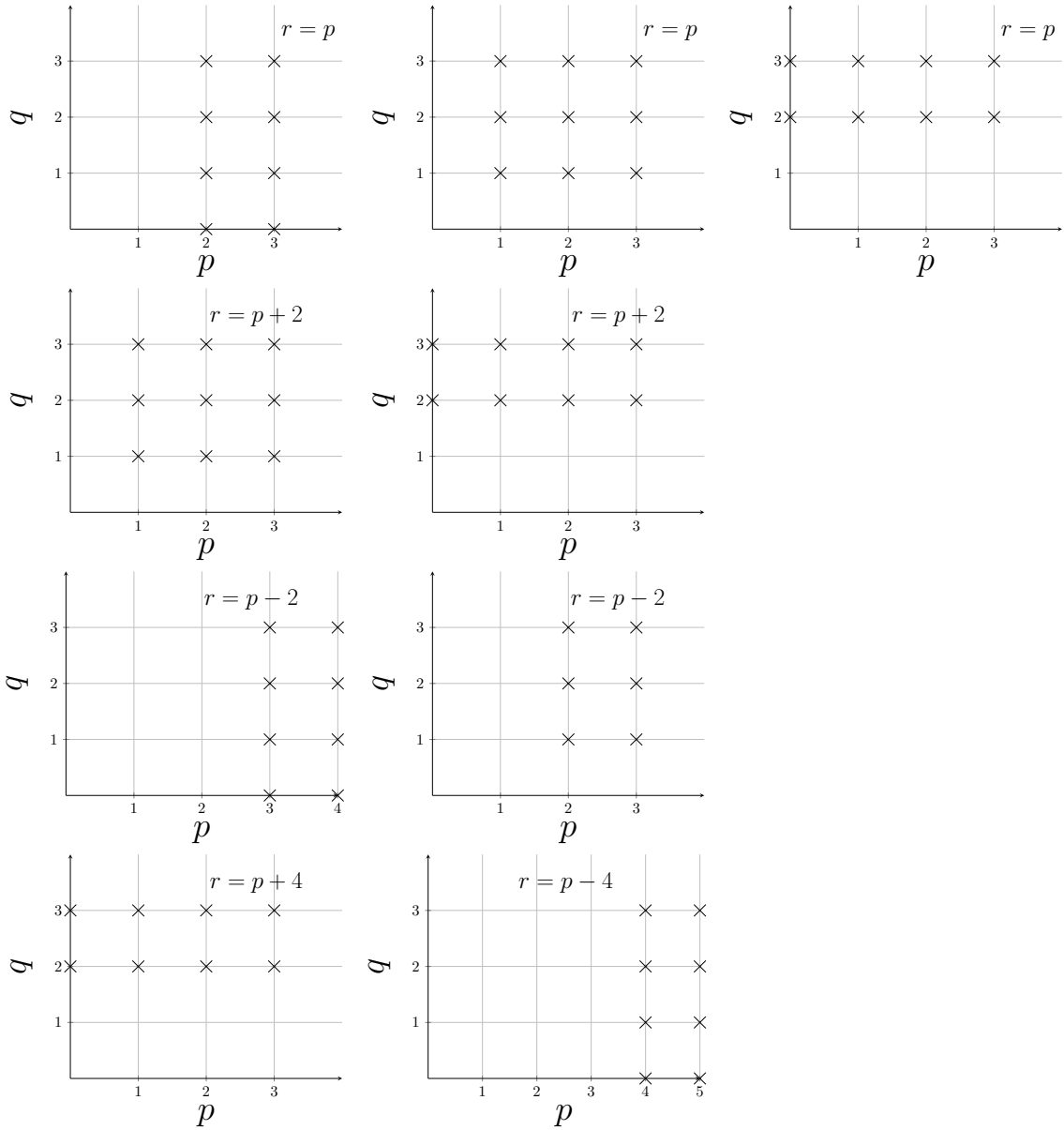
\begin{figure}[H]
    \centering
    \begin{tabular}{c c c c}
\begin{tikzpicture}
\begin{axis}[%
    axis lines = left,
    xmin = 0, xmax = 4,
    ymin = 0, ymax = 4,
    xlabel = {\Huge $p$}, ylabel = {\Huge $q$},
    xtick={1,2,3}, ytick={1,2,3},
    xmajorgrids=true, ymajorgrids=true,
]
\addplot [only marks, mark=x, mark size=6pt] table {
2   0
2   1
2   2
2   3
3   0
3   1
3   2
3   3
};
\node at (axis cs:3.5,3.5) [anchor=center] {\LARGE $r=p$};
\end{axis}
\end{tikzpicture}
&
\begin{tikzpicture}
\begin{axis}[%
    axis lines = left,
    xmin = 0, xmax = 4,
    ymin = 0, ymax = 4,
    xlabel = {\Huge $p$}, ylabel = {\Huge $q$},
    xtick={1,2,3}, ytick={1,2,3},
    xmajorgrids=true, ymajorgrids=true,
]
\addplot [only marks, mark=x, mark size=6pt] table {
1   1
1   2
1   3
2   1
2   2
2   3
3   1
3   2
3   3
};
\node at (axis cs:3.5,3.5) [anchor=center] {\LARGE $r=p$};
\end{axis}
\end{tikzpicture}
&
\begin{tikzpicture}
\begin{axis}[%
    axis lines = left,
    xmin = 0, xmax = 4,
    ymin = 0, ymax = 4,
    xlabel = {\Huge $p$}, ylabel = {\Huge $q$},
    xtick={1,2,3}, ytick={1,2,3},
    xmajorgrids=true, ymajorgrids=true,
]
\addplot [only marks, mark=x, mark size=6pt] table {
0   2
0   3
1   2
1   3
2   2
2   3
3   2
3   3
};
\node at (axis cs:3.5,3.5) [anchor=center] {\LARGE $r=p$};
\end{axis}
\end{tikzpicture}
\\
\begin{tikzpicture}
\begin{axis}[%
    axis lines = left,
    xmin = 0, xmax = 4,
    ymin = 0, ymax = 4,
    xlabel = {\Huge $p$}, ylabel = {\Huge $q$},
    xtick={1,2,3}, ytick={1,2,3},
    xmajorgrids=true, ymajorgrids=true,
]
\addplot [only marks, mark=x, mark size=6pt] table {
1   1
1   2
1   3
2   1
2   2
2   3
3   1
3   2
3   3
};
\node at (axis cs:3.5,3.5) [anchor=east] {\LARGE $r=p+2$};
\end{axis}
\end{tikzpicture}
&
\begin{tikzpicture}
\begin{axis}[%
    axis lines = left,
    xmin = 0, xmax = 4,
    ymin = 0, ymax = 4,
    xlabel = {\Huge $p$}, ylabel = {\Huge $q$},
    xtick={1,2,3}, ytick={1,2,3},
    xmajorgrids=true, ymajorgrids=true,
]
\addplot [only marks, mark=x, mark size=6pt] table {
0   2
0   3
1   2
1   3
2   2
2   3
3   2
3   3
};
\node at (axis cs:3.5,3.5) [anchor=east] {\LARGE $r=p+2$};
\end{axis}
\end{tikzpicture}
\\
\begin{tikzpicture}
\begin{axis}[%
    axis lines = left,
    xmin = 0, xmax = 4,
    ymin = 0, ymax = 4,
    xlabel = {\Huge $p$}, ylabel = {\Huge $q$},
    xtick={1,2,3,4}, ytick={1,2,3},
    xmajorgrids=true, ymajorgrids=true,
]
\addplot [only marks, mark=x, mark size=6pt] table {
3   0
3   1
3   2
3   3
4   0
4   1
4   2
4   3
};
\node at (axis cs:3.5,3.5) [anchor=east] {\LARGE $r=p-2$};
\end{axis}
\end{tikzpicture}
&
\begin{tikzpicture}
\begin{axis}[%
    axis lines = left,
    xmin = 0, xmax = 4,
    ymin = 0, ymax = 4,
    xlabel = {\Huge $p$}, ylabel = {\Huge $q$},
    xtick={1,2,3}, ytick={1,2,3},
    xmajorgrids=true, ymajorgrids=true,
]
\addplot [only marks, mark=x, mark size=6pt] table {
2   1
2   2
2   3
3   1
3   2
3   3
};
\node at (axis cs:3.5,3.5) [anchor=east] {\LARGE $r=p-2$};
\end{axis}
\end{tikzpicture}
\\
\begin{tikzpicture}
\begin{axis}[%
    axis lines = left,
    xmin = 0, xmax = 4,
    ymin = 0, ymax = 4,
    xlabel = {\Huge $p$}, ylabel = {\Huge $q$},
    xtick={1,2,3}, ytick={1,2,3},
    xmajorgrids=true, ymajorgrids=true,
]
\addplot [only marks, mark=x, mark size=6pt] table {
0   2
0   3
1   2
1   3
2   2
2   3
3   2
3   3
};
\node at (axis cs:3.5,3.5) [anchor=east] {\LARGE $r=p+4$};
\end{axis}
\end{tikzpicture}
&
\begin{tikzpicture}
\begin{axis}[%
    axis lines = left,
    xmin = 0, xmax = 5,
    ymin = 0, ymax = 4,
    xlabel = {\Huge $p$}, ylabel = {\Huge $q$},
    xtick={1,2,3,4,5}, ytick={1,2,3},
    xmajorgrids=true, ymajorgrids=true,
]
\addplot [only marks, mark=x, mark size=6pt] table {
4   0
4   1
4   2
4   3
5   0
5   1
5   2
5   3
};
\node at (axis cs:3.5,3.5) [anchor=east] {\LARGE $r=p-4$};
\end{axis}
\end{tikzpicture}
\end{tabular}%
    \caption{(2,2) towers}
    \label{fig:my_label}
\end{figure}

\section{$SO(7)$ tensor tower diagrams}

\begin{figure}[H]
\centering
\begin{tikzpicture}
\begin{axis}[
    axis lines = left,
    xmin = 0, xmax = 4,
    ymin = 0, ymax = 4,
    xlabel = {\Huge $p$}, ylabel = {\Huge $q$},
    xtick={1,2,3}, ytick={1,2,3},
    xmajorgrids=true, ymajorgrids=true,
]
\addplot [only marks, mark=x, mark size=6pt] table {
0   0
0   1
0   2
0   3
1   0
1   1
1   2
1   3
2   0
2   1
2   2
2   3
3   0
3   1
3   2
3   3
};
\node at (axis cs:3.5,3.5) [anchor=center] {\LARGE $r=p$};
\end{axis}
\end{tikzpicture}%
\caption{Scalar  tower.}
\end{figure}
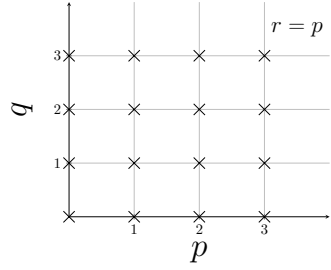

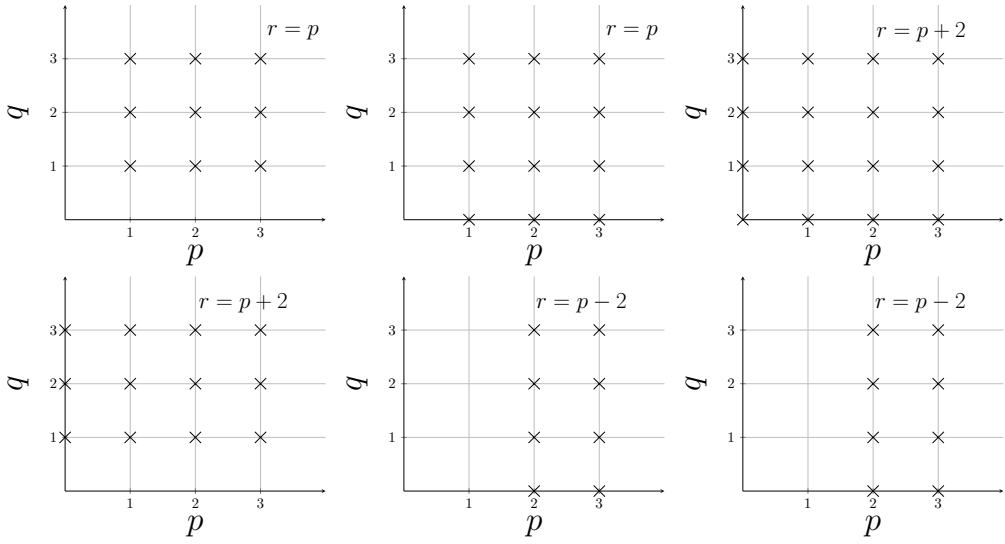
\begin{figure}[H]
\centering
\begin{tabular}{l l l l l l}
\begin{tikzpicture}
\begin{axis}[%
    axis lines = left,
    xmin = 0, xmax = 4,
    ymin = 0, ymax = 4,
    xlabel = {\Huge $p$}, ylabel = {\Huge $q$},
    xtick={1,2,3}, ytick={1,2,3},
    xmajorgrids=true, ymajorgrids=true,
]
\addplot [only marks, mark=x, mark size=6pt] table {
1   1
1   2
1   3
2   1
2   2
2   3
3   1
3   2
3   3
};
\node at (axis cs:3.5,3.5) [anchor=center] {\LARGE $r=p$};
\end{axis}
\end{tikzpicture} 
&
\begin{tikzpicture}
\begin{axis}[%
    axis lines = left,
    xmin = 0, xmax = 4,
    ymin = 0, ymax = 4,
    xlabel = {\Huge $p$}, ylabel = {\Huge $q$},
    xtick={1,2,3}, ytick={1,2,3},
    xmajorgrids=true, ymajorgrids=true,
]
\addplot [only marks, mark=x, mark size=6pt] table {
1   0
1   1
1   2
1   3
2   0
2   1
2   2
2   3
3   0
3   1
3   2
3   3
};
\node at (axis cs:3.5,3.5) [anchor=center] {\LARGE $r=p$};
\end{axis}
\end{tikzpicture}
&
\begin{tikzpicture}
\begin{axis}[%
    axis lines = left,
    xmin = 0, xmax = 4,
    ymin = 0, ymax = 4,
    xlabel = {\Huge $p$}, ylabel = {\Huge $q$},
    xtick={1,2,3}, ytick={1,2,3},
    xmajorgrids=true, ymajorgrids=true,
]
\addplot [only marks, mark=x, mark size=6pt] table {
0   0
0   1
0   2
0   3
1   0
1   1
1   2
1   3
2   0
2   1
2   2
2   3
3   0
3   1
3   2
3   3
};
\node at (axis cs:3.5,3.5) [anchor=east] {\LARGE $r=p+2$};
\end{axis}
\end{tikzpicture}
\\
\begin{tikzpicture}
\begin{axis}[%
    axis lines = left,
    xmin = 0, xmax = 4,
    ymin = 0, ymax = 4,
    xlabel = {\Huge $p$}, ylabel = {\Huge $q$},
    xtick={1,2,3}, ytick={1,2,3},
    xmajorgrids=true, ymajorgrids=true,
]
\addplot [only marks, mark=x, mark size=6pt] table {
0   1
0   2
0   3
1   1
1   2
1   3
2   1
2   2
2   3
3   1
3   2
3   3
};
\node at (axis cs:3.5,3.5) [anchor=east] {\LARGE $r=p+2$};
\end{axis}
\end{tikzpicture}
&
\begin{tikzpicture}
\begin{axis}[%
    axis lines = left,
    xmin = 0, xmax = 4,
    ymin = 0, ymax = 4,
    xlabel = {\Huge $p$}, ylabel = {\Huge $q$},
    xtick={1,2,3}, ytick={1,2,3},
    xmajorgrids=true, ymajorgrids=true,
]
\addplot [only marks, mark=x, mark size=6pt] table {
2   0
2   1
2   2
2   3
3   0
3   1
3   2
3   3
};
\node at (axis cs:3.5,3.5) [anchor=east] {\LARGE $r=p-2$};
\end{axis}
\end{tikzpicture}
&
\begin{tikzpicture}
\begin{axis}[%
    axis lines = left,
    xmin = 0, xmax = 4,
    ymin = 0, ymax = 4,
    xlabel = {\Huge $p$}, ylabel = {\Huge $q$},
    xtick={1,2,3}, ytick={1,2,3},
    xmajorgrids=true, ymajorgrids=true,
]
\addplot [only marks, mark=x, mark size=6pt] table {
2   0
2   1
2   2
2   3
3   0
3   1
3   2
3   3
};
\node at (axis cs:3.5,3.5) [anchor=east] {\LARGE $r=p-2$};
\end{axis}%
\end{tikzpicture}%
\end{tabular}%
\caption{One-form towers.}
\end{figure}

\begin{figure}[H]
\centering
\begin{tabular}{l l l l l l l l}
\begin{tikzpicture}%
\begin{axis}[%
    axis lines = left,
    xmin = 0, xmax = 4,
    ymin = 0, ymax = 4,
    xlabel = {\Huge $p$}, ylabel = {\Huge $q$},
    xtick={1,2,3}, ytick={1,2,3},
    xmajorgrids=true, ymajorgrids=true,
]
\addplot [only marks, mark=x, mark size=6pt] table {
0   0
0   1
0   2
0   3
1   0
1   1
1   2
1   3
2   0
2   1
2   2
2   3
3   0
3   1
3   2
3   3
};
\node at (axis cs:3.5,3.5) [anchor=center] {\LARGE $r=p$};
\end{axis}
\end{tikzpicture} 
&
\begin{tikzpicture}
\begin{axis}[%
    axis lines = left,
    xmin = 0, xmax = 4,
    ymin = 0, ymax = 4,
    xlabel = {\Huge $p$}, ylabel = {\Huge $q$},
    xtick={1,2,3}, ytick={1,2,3},
    xmajorgrids=true, ymajorgrids=true,
]
\addplot [only marks, mark=x, mark size=6pt] table {
0   1
0   2
0   3
1   0
1   1
1   2
1   3
2   0
2   1
2   2
2   3
3   0
3   1
3   2
3   3
};
\node at (axis cs:3.5,3.5) [anchor=center] {\LARGE $r=p$};
\end{axis}
\end{tikzpicture}
&
\begin{tikzpicture}
\begin{axis}[%
    axis lines = left,
    xmin = 0, xmax = 4,
    ymin = 0, ymax = 4,
    xlabel = {\Huge $p$}, ylabel = {\Huge $q$},
    xtick={1,2,3}, ytick={1,2,3},
    xmajorgrids=true, ymajorgrids=true,
]
\addplot [only marks, mark=x, mark size=6pt] table {
1   1
1   2
1   3
2   1
2   2
2   3
3   1
3   2
3   3
};
\node at (axis cs:3.5,3.5) [anchor=center] {\LARGE $r=p$};
\end{axis}
\end{tikzpicture}
&
\begin{tikzpicture}
\begin{axis}[%
    axis lines = left,
    xmin = 0, xmax = 4,
    ymin = 0, ymax = 4,
    xlabel = {\Huge $p$}, ylabel = {\Huge $q$},
    xtick={1,2,3}, ytick={1,2,3},
    xmajorgrids=true, ymajorgrids=true,
]
\addplot [only marks, mark=x, mark size=6pt] table {
1   0
1   1
1   2
1   3
2   0
2   1
2   2
2   3
3   0
3   1
3   2
3   3
};
\node at (axis cs:3.5,3.5) [anchor=center] {\LARGE $r=p$};
\end{axis}
\end{tikzpicture}
\\
\begin{tikzpicture}
\begin{axis}[%
    axis lines = left,
    xmin = 0, xmax = 4,
    ymin = 0, ymax = 4,
    xlabel = {\Huge $p$}, ylabel = {\Huge $q$},
    xtick={1,2,3}, ytick={1,2,3},
    xmajorgrids=true, ymajorgrids=true,
]
\addplot [only marks, mark=x, mark size=6pt] table {
0   0
0   1
0   2
0   3
1   0
1   1
1   2
1   3
2   0
2   1
2   2
2   3
3   0
3   1
3   2
3   3
};
\node at (axis cs:3.5,3.5) [anchor=east] {\LARGE $r=p+2$};
\end{axis}
\end{tikzpicture}
&
\begin{tikzpicture}
\begin{axis}[%
    axis lines = left,
    xmin = 0, xmax = 4,
    ymin = 0, ymax = 4,
    xlabel = {\Huge $p$}, ylabel = {\Huge $q$},
    xtick={1,2,3}, ytick={1,2,3},
    xmajorgrids=true, ymajorgrids=true,
]
\addplot [only marks, mark=x, mark size=6pt] table {
0   1
0   2
0   3
1   1
1   2
1   3
2   1
2   2
2   3
3   1
3   2
3   3
};
\node at (axis cs:3.5,3.5) [anchor=east] {\LARGE $r=p+2$};
\end{axis}
\end{tikzpicture}
&
\begin{tikzpicture}
\begin{axis}[%
    axis lines = left,
    xmin = 0, xmax = 4,
    ymin = 0, ymax = 4,
    xlabel = {\Huge $p$}, ylabel = {\Huge $q$},
    xtick={1,2,3}, ytick={1,2,3},
    xmajorgrids=true, ymajorgrids=true,
]
\addplot [only marks, mark=x, mark size=6pt] table {
2   0
2   1
2   2
2   3
3   0
3   1
3   2
3   3
};
\node at (axis cs:3.5,3.5) [anchor=east] {\LARGE $r=p-2$};
\end{axis}
\end{tikzpicture}
&
\begin{tikzpicture}
\begin{axis}[%
    axis lines = left,
    xmin = 0, xmax = 4,
    ymin = 0, ymax = 4,
    xlabel = {\Huge $p$}, ylabel = {\Huge $q$},
    xtick={1,2,3}, ytick={1,2,3},
    xmajorgrids=true, ymajorgrids=true,
]
\addplot [only marks, mark=x, mark size=6pt] table {
2   0
2   1
2   2
2   3
3   0
3   1
3   2
3   3
};
\node at (axis cs:3.5,3.5) [anchor=east] {\LARGE $r=p-2$};
\end{axis}%
\end{tikzpicture}%
\end{tabular}%
\caption{ $\slashed{D}_{1/2}$ towers.}
\label{fig:apx:D12towers}
\end{figure}

\begin{figure}[H]
\centering
\begin{tabular}{l l l l}
\begin{tikzpicture}%
\begin{axis}[%
    axis lines = left,
    xmin = 0, xmax = 4,
    ymin = 0, ymax = 4,
    xlabel = {\Huge $p$}, ylabel = {\Huge $q$},
    xtick={1,2,3}, ytick={1,2,3},
    xmajorgrids=true, ymajorgrids=true,
]
\addplot [only marks, mark=x, mark size=6pt] table {
1   1
1   2
1   3
2   1
2   2
2   3
3   1
3   2
3   3
};
\node at (axis cs:3.5,3.5) [anchor=center] {\LARGE $r=p$};
\end{axis}
\end{tikzpicture} 
&
\begin{tikzpicture}
\begin{axis}[%
    axis lines = left,
    xmin = 0, xmax = 4,
    ymin = 0, ymax = 4,
    xlabel = {\Huge $p$}, ylabel = {\Huge $q$},
    xtick={1,2,3}, ytick={1,2,3},
    xmajorgrids=true, ymajorgrids=true,
]
\addplot [only marks, mark=x, mark size=6pt] table {
0   1
0   2
0   3
1   1
1   2
1   3
2   1
2   2
2   3
3   1
3   2
3   3
};
\node at (axis cs:3.5,3.5) [anchor=center] {\LARGE $r=p$};
\end{axis}
\end{tikzpicture}
&
\begin{tikzpicture}
\begin{axis}[%
    axis lines = left,
    xmin = 0, xmax = 4,
    ymin = 0, ymax = 4,
    xlabel = {\Huge $p$}, ylabel = {\Huge $q$},
    xtick={1,2,3}, ytick={1,2,3},
    xmajorgrids=true, ymajorgrids=true,
]
\addplot [only marks, mark=x, mark size=6pt] table {
1   0
1   1
1   2
1   3
2   0
2   1
2   2
2   3
3   0
3   1
3   2
3   3
};
\node at (axis cs:3.5,3.5) [anchor=center] {\LARGE $r=p$};
\end{axis}
\end{tikzpicture}
&
\begin{tikzpicture}
\begin{axis}[%
    axis lines = left,
    xmin = 0, xmax = 4,
    ymin = 0, ymax = 4,
    xlabel = {\Huge $p$}, ylabel = {\Huge $q$},
    xtick={1,2,3}, ytick={1,2,3},
    xmajorgrids=true, ymajorgrids=true,
]
\addplot [only marks, mark=x, mark size=6pt] table {
1   0
1   1
1   2
1   3
2   0
2   1
2   2
2   3
3   0
3   1
3   2
3   3
};
\node at (axis cs:3.5,3.5) [anchor=center] {\LARGE $r=p$};
\end{axis}
\end{tikzpicture}
\\
\begin{tikzpicture}
\begin{axis}[%
    axis lines = left,
    xmin = 0, xmax = 4,
    ymin = 0, ymax = 4,
    xlabel = {\Huge $p$}, ylabel = {\Huge $q$},
    xtick={1,2,3}, ytick={1,2,3},
    xmajorgrids=true, ymajorgrids=true,
]
\addplot [only marks, mark=x, mark size=6pt] table {
2   0
2   1
2   2
2   3
3   0
3   1
3   2
3   3
};
\node at (axis cs:3.5,3.5) [anchor=center] {\LARGE $r=p$};
\end{axis}
\end{tikzpicture}
&
\begin{tikzpicture}
\begin{axis}[%
    axis lines = left,
    xmin = 0, xmax = 4,
    ymin = 0, ymax = 4,
    xlabel = {\Huge $p$}, ylabel = {\Huge $q$},
    xtick={1,2,3}, ytick={1,2,3},
    xmajorgrids=true, ymajorgrids=true,
]
\addplot [only marks, mark=x, mark size=6pt] table {
0   0
0   1
0   2
0   3
1   0
1   1
1   2
1   3
2   0
2   1
2   2
2   3
3   0
3   1
3   2
3   3
};
\node at (axis cs:3.5,3.5) [anchor=east] {\LARGE $r=p+2$};
\end{axis}
\end{tikzpicture}
&
\begin{tikzpicture}
\begin{axis}[%
    axis lines = left,
    xmin = 0, xmax = 4,
    ymin = 0, ymax = 4,
    xlabel = {\Huge $p$}, ylabel = {\Huge $q$},
    xtick={1,2,3}, ytick={1,2,3},
    xmajorgrids=true, ymajorgrids=true,
]
\addplot [only marks, mark=x, mark size=6pt] table {
0   1
0   2
0   3
1   1
1   2
1   3
2   1
2   2
2   3
3   1
3   2
3   3
};
\node at (axis cs:3.5,3.5) [anchor=east] {\LARGE $r=p+2$};
\end{axis}
\end{tikzpicture}
&
\begin{tikzpicture}
\begin{axis}[%
    axis lines = left,
    xmin = 0, xmax = 4,
    ymin = 0, ymax = 4,
    xlabel = {\Huge $p$}, ylabel = {\Huge $q$},
    xtick={1,2,3}, ytick={1,2,3},
    xmajorgrids=true, ymajorgrids=true,
]
\addplot [only marks, mark=x, mark size=6pt] table {
1   0
1   1
1   2
1   3
2   0
2   1
2   2
2   3
3   0
3   1
3   2
3   3
};
\node at (axis cs:3.5,3.5) [anchor=east] {\LARGE $r=p+2$};
\end{axis}
\end{tikzpicture}
\\
\begin{tikzpicture}
\begin{axis}[%
    axis lines = left,
    xmin = 0, xmax = 4,
    ymin = 0, ymax = 4,
    xlabel = {\Huge $p$}, ylabel = {\Huge $q$},
    xtick={1,2,3}, ytick={1,2,3},
    xmajorgrids=true, ymajorgrids=true,
]
\addplot [only marks, mark=x, mark size=6pt] table {
0   2
0   3
1   2
1   3
2   2
2   3
3   2
3   3
};
\node at (axis cs:3.5,3.5) [anchor=east] {\LARGE $r=p+2$};
\end{axis}
\end{tikzpicture}
&
\begin{tikzpicture}
\begin{axis}[%
    axis lines = left,
    xmin = 0, xmax = 4,
    ymin = 0, ymax = 4,
    xlabel = {\Huge $p$}, ylabel = {\Huge $q$},
    xtick={1,2,3}, ytick={1,2,3},
    xmajorgrids=true, ymajorgrids=true,
]
\addplot [only marks, mark=x, mark size=6pt] table {
2   0
2   1
2   2
2   3
3   0
3   1
3   2
3   3
};
\node at (axis cs:3.5,3.5) [anchor=east] {\LARGE $r=p-2$};
\end{axis}
\end{tikzpicture}
&
\begin{tikzpicture}
\begin{axis}[%
    axis lines = left,
    xmin = 0, xmax = 4,
    ymin = 0, ymax = 4,
    xlabel = {\Huge $p$}, ylabel = {\Huge $q$},
    xtick={1,2,3}, ytick={1,2,3},
    xmajorgrids=true, ymajorgrids=true,
]
\addplot [only marks, mark=x, mark size=6pt] table {
2   0
2   1
2   2
2   3
3   0
3   1
3   2
3   3
};
\node at (axis cs:3.5,3.5) [anchor=east] {\LARGE $r=p-2$};
\end{axis}
\end{tikzpicture}
&
\begin{tikzpicture}
\begin{axis}[%
    axis lines = left,
    xmin = 0, xmax = 4,
    ymin = 0, ymax = 4,
    xlabel = {\Huge $p$}, ylabel = {\Huge $q$},
    xtick={1,2,3}, ytick={1,2,3},
    xmajorgrids=true, ymajorgrids=true,
]
\addplot [only marks, mark=x, mark size=6pt] table {
2   1
2   2
2   3
3   1
3   2
3   3
};
\node at (axis cs:3.5,3.5) [anchor=east] {\LARGE $r=p-2$};
\end{axis}
\end{tikzpicture}
\\
\begin{tikzpicture}
\begin{axis}[%
    axis lines = left,
    xmin = 0, xmax = 5,
    ymin = 0, ymax = 4,
    xlabel = {\Huge $p$}, ylabel = {\Huge $q$},
    xtick={1,2,3,4}, ytick={1,2,3},
    xmajorgrids=true, ymajorgrids=true,
]
\addplot [only marks, mark=x, mark size=6pt] table {
3   0
3   1
3   2
3   3
4   0
4   1
4   2
4   3
};
\node at (axis cs:3.5,3.5) [anchor=east] {\LARGE $r=p-2$};
\end{axis}
\end{tikzpicture}
&
\begin{tikzpicture}
\begin{axis}[%
    axis lines = left,
    xmin = 0, xmax = 4,
    ymin = 0, ymax = 4,
    xlabel = {\Huge $p$}, ylabel = {\Huge $q$},
    xtick={1,2,3}, ytick={1,2,3},
    xmajorgrids=true, ymajorgrids=true,
]
\addplot [only marks, mark=x, mark size=6pt] table {
0   1
0   2
0   3
1   1
1   2
1   3
2   1
2   2
2   3
3   1
3   2
3   3
};
\node at (axis cs:3.5,3.5) [anchor=east] {\LARGE $r=p+4$};
\end{axis}
\end{tikzpicture}
&
\begin{tikzpicture}
\begin{axis}[%
    axis lines = left,
    xmin = 0, xmax = 6,
    ymin = 0, ymax = 4,
    xlabel = {\Huge $p$}, ylabel = {\Huge $q$},
    xtick={1,2,3,4,5}, ytick={1,2,3},
    xmajorgrids=true, ymajorgrids=true,
]
\addplot [only marks, mark=x, mark size=6pt] table {
4   0
4   1
4   2
4   3
5   0
5   1
5   2
5   3
};
\node at (axis cs:3.5,3.5) [anchor=east] {\LARGE $r=p-4$};
\end{axis}
\end{tikzpicture}%
\end{tabular}%
\caption{Two-form towers.}
\label{fig:apx:2formtowers}
\end{figure}
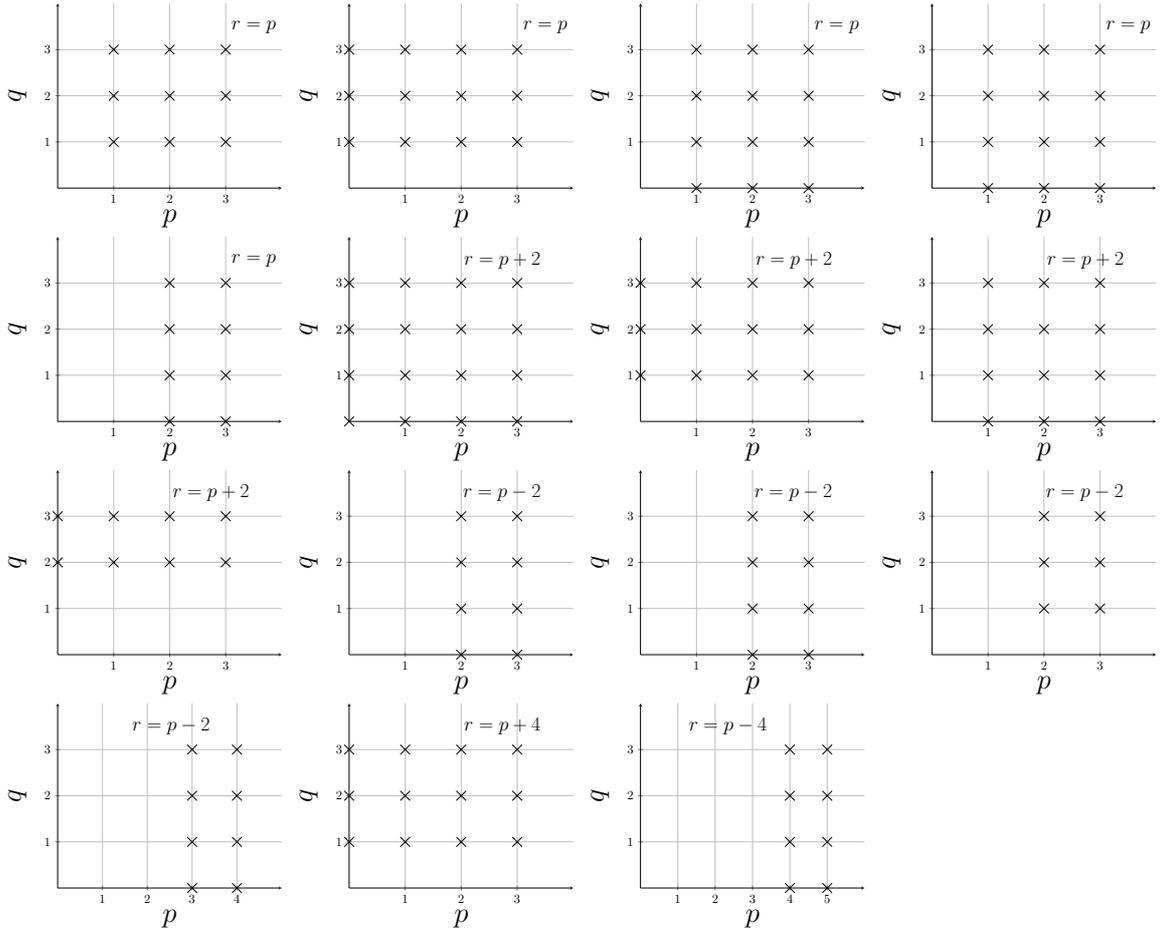

\begin{figure}[H]
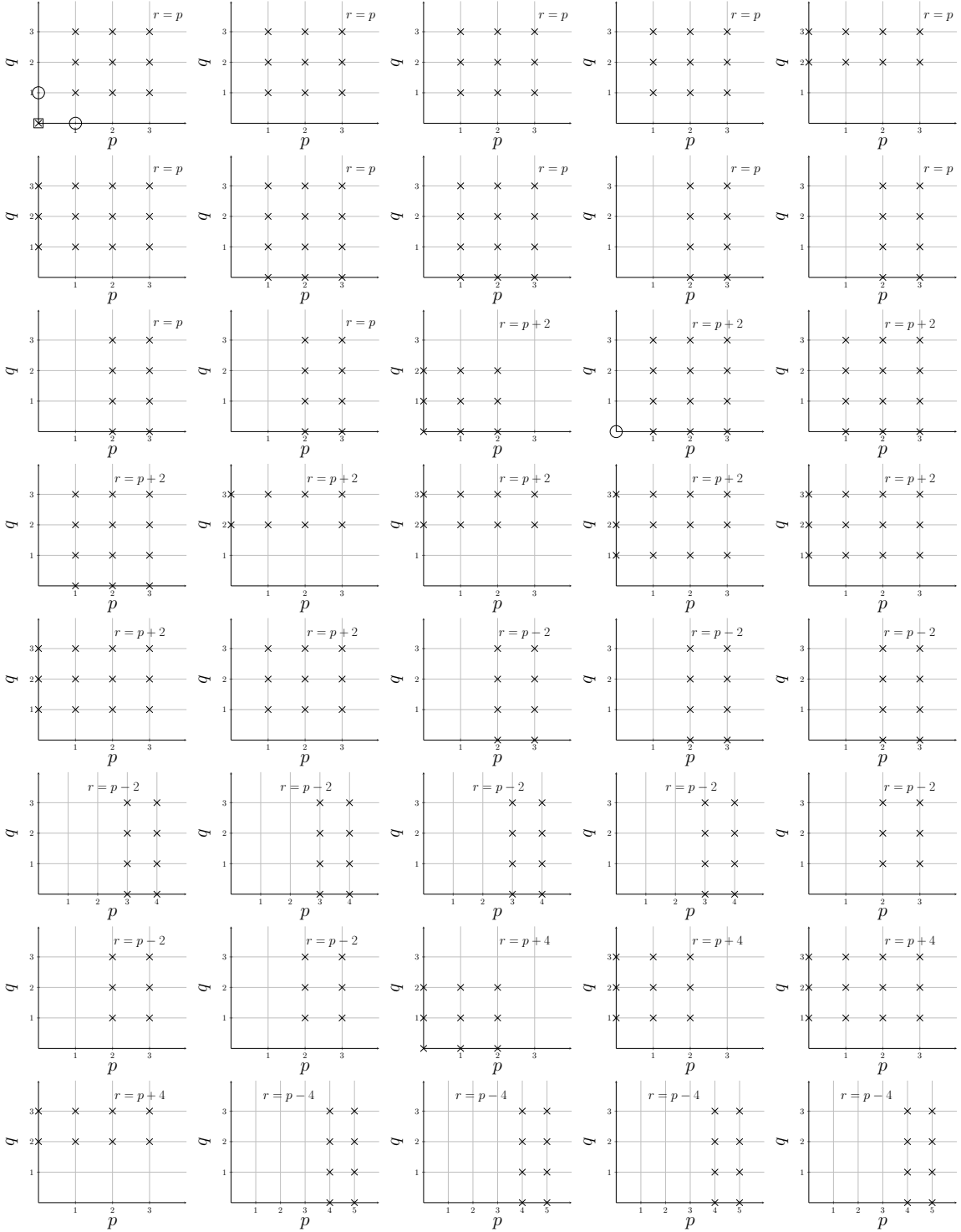

\centering
\includestandalone[scale=0.34]{tikz(NPP)/D32towers}
\caption{ $\slashed{D}_{3/2}$ towers. The empty circles mark modes which appear from the decomposition of representations of $SO(8)$ on the round sphere, but do not exist when derived in the conventional method. The excess modes are $(1,0;1)=({\bf 4,2})$, $(0,1;0)=({\bf 5,1})$ and $(0,0;2)=({\bf 1,3})$. The square with a cross, on the other hand, emphasises the fact that the $G$  irrep $(0,0;0)=({\bf 1,1)}$  appears in the squashed spectrum but is not produced 
in the spontaneous symmetry breaking of the round sphere spectrum.}
\label{fig:apx:D32towers}
\end{figure}

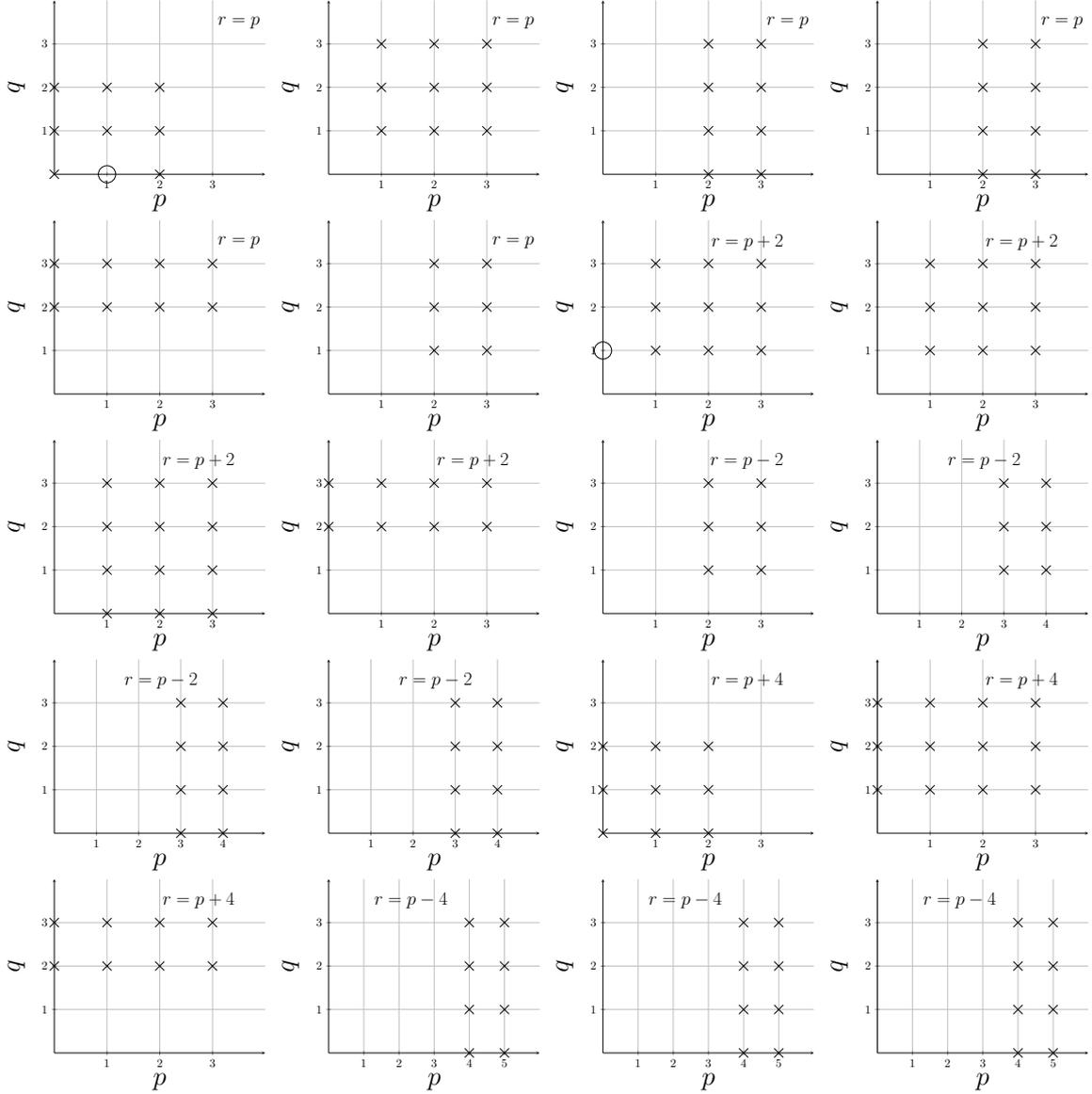
\begin{figure}[H]
\centering
\begin{tabular}{l l l l l l l l}
\begin{tikzpicture}%
\begin{axis}[%
    axis lines = left,
    xmin = 0, xmax = 4,
    ymin = 0, ymax = 4,
    xlabel = {\Huge $p$}, ylabel = {\Huge $q$},
    xtick={1,2,3}, ytick={1,2,3},
    xmajorgrids=true, ymajorgrids=true,
]
\addplot [only marks, mark=x, mark size=6pt] table {
0	0
0   1
0	2
1	1
1	2
2   0
2   1
2   2
};
\addplot [only marks, mark=o, mark size=8pt] table {
1   0
};
\node at (axis cs:3.5,3.5) [anchor=center] {\LARGE $r=p$};
\end{axis}
\end{tikzpicture} 
&
\begin{tikzpicture}%
\begin{axis}[%
    axis lines = left,
    xmin = 0, xmax = 4,
    ymin = 0, ymax = 4,
    xlabel = {\Huge $p$}, ylabel = {\Huge $q$},
    xtick={1,2,3}, ytick={1,2,3},
    xmajorgrids=true, ymajorgrids=true,
]
\addplot [only marks, mark=x, mark size=6pt] table {
1   1
1   2
1   3
2   1
2   2
2   3
3   1
3   2
3   3
};
\node at (axis cs:3.5,3.5) [anchor=center] {\LARGE $r=p$};
\end{axis}
\end{tikzpicture} 
&
\begin{tikzpicture}%
\begin{axis}[%
    axis lines = left,
    xmin = 0, xmax = 4,
    ymin = 0, ymax = 4,
    xlabel = {\Huge $p$}, ylabel = {\Huge $q$},
    xtick={1,2,3}, ytick={1,2,3},
    xmajorgrids=true, ymajorgrids=true,
]
\addplot [only marks, mark=x, mark size=6pt] table {
2	0
2	1
2	2
2   3
3	0
3	1
3	2
3	3
};
\node at (axis cs:3.5,3.5) [anchor=center] {\LARGE $r=p$};
\end{axis}
\end{tikzpicture} 
&
\begin{tikzpicture}%
\begin{axis}[%
    axis lines = left,
    xmin = 0, xmax = 4,
    ymin = 0, ymax = 4,
    xlabel = {\Huge $p$}, ylabel = {\Huge $q$},
    xtick={1,2,3}, ytick={1,2,3},
    xmajorgrids=true, ymajorgrids=true,
]
\addplot [only marks, mark=x, mark size=6pt] table {
2	0
2	1
2	2
2   3
3	0
3	1
3	2
3	3
};
\node at (axis cs:3.5,3.5) [anchor=center] {\LARGE $r=p$};
\end{axis}
\end{tikzpicture} 
\\
\begin{tikzpicture}
\begin{axis}[%
    axis lines = left,
    xmin = 0, xmax = 4,
    ymin = 0, ymax = 4,
    xlabel = {\Huge $p$}, ylabel = {\Huge $q$},
    xtick={1,2,3}, ytick={1,2,3},
    xmajorgrids=true, ymajorgrids=true,
]
\addplot [only marks, mark=x, mark size=6pt] table {
0	2
0	3
1	2
1	3
2	2
2	3
3	2
3	3
};
\node at (axis cs:3.5,3.5) [anchor=center] {\LARGE $r=p$};
\end{axis}
\end{tikzpicture}
&
\begin{tikzpicture}
\begin{axis}[%
    axis lines = left,
    xmin = 0, xmax = 4,
    ymin = 0, ymax = 4,
    xlabel = {\Huge $p$}, ylabel = {\Huge $q$},
    xtick={1,2,3}, ytick={1,2,3},
    xmajorgrids=true, ymajorgrids=true,
]
\addplot [only marks, mark=x, mark size=6pt] table {
2   1
2   2
2   3
3   1
3   2
3   3
};
\node at (axis cs:3.5,3.5) [anchor=center] {\LARGE $r=p$};
\end{axis}
\end{tikzpicture}
&
\begin{tikzpicture}%
\begin{axis}[%
    axis lines = left,
    xmin = 0, xmax = 4,
    ymin = 0, ymax = 4,
    xlabel = {\Huge $p$}, ylabel = {\Huge $q$},
    xtick={1,2,3}, ytick={1,2,3},
    xmajorgrids=true, ymajorgrids=true,
]
\addplot [only marks, mark=x, mark size=6pt] table {
1   1
1   2
1   3
2   1
2   2
2   3
3   1
3   2
3   3
};
\addplot [only marks, mark=o, mark size=8pt] table {
0   1
};
\node at (axis cs:3.5,3.5) [anchor=east] {\LARGE $r=p+2$};
\end{axis}
\end{tikzpicture} 
&
\begin{tikzpicture}%
\begin{axis}[%
    axis lines = left,
    xmin = 0, xmax = 4,
    ymin = 0, ymax = 4,
    xlabel = {\Huge $p$}, ylabel = {\Huge $q$},
    xtick={1,2,3}, ytick={1,2,3},
    xmajorgrids=true, ymajorgrids=true,
]
\addplot [only marks, mark=x, mark size=6pt] table {
1   1
1   2
1   3
2   1
2   2
2   3
3   1
3   2
3   3
};

\node at (axis cs:3.5,3.5) [anchor=east] {\LARGE $r=p+2$};
\end{axis}
\end{tikzpicture} 
\\
\begin{tikzpicture}%
\begin{axis}[%
    axis lines = left,
    xmin = 0, xmax = 4,
    ymin = 0, ymax = 4,
    xlabel = {\Huge $p$}, ylabel = {\Huge $q$},
    xtick={1,2,3}, ytick={1,2,3},
    xmajorgrids=true, ymajorgrids=true,
]
\addplot [only marks, mark=x, mark size=6pt] table {
1	0
1	1
1	2
1	3
2	0
2	1
2	2
2	3
3   0
3   1
3   2
3   3
};
\node at (axis cs:3.5,3.5) [anchor=east] {\LARGE $r=p+2$};
\end{axis}
\end{tikzpicture} 
&
\begin{tikzpicture}%
\begin{axis}[%
    axis lines = left,
    xmin = 0, xmax = 4,
    ymin = 0, ymax = 4,
    xlabel = {\Huge $p$}, ylabel = {\Huge $q$},
    xtick={1,2,3}, ytick={1,2,3},
    xmajorgrids=true, ymajorgrids=true,
]
\addplot [only marks, mark=x, mark size=6pt] table {
0	2
0	3
1	2
1	3
2	2
2	3
3	2
3	3
};
\node at (axis cs:3.5,3.5) [anchor=east] {\LARGE $r=p+2$};
\end{axis}
\end{tikzpicture}
&
\begin{tikzpicture}%
\begin{axis}[%
    axis lines = left,
    xmin = 0, xmax = 4,
    ymin = 0, ymax = 4,
    xlabel = {\Huge $p$}, ylabel = {\Huge $q$},
    xtick={1,2,3}, ytick={1,2,3},
    xmajorgrids=true, ymajorgrids=true,
]
\addplot [only marks, mark=x, mark size=6pt] table {
2   1
2   2
2   3
3   1
3   2
3   3
};
\node at (axis cs:3.5,3.5) [anchor=east] {\LARGE $r=p-2$};
\end{axis}
\end{tikzpicture} 
&
\begin{tikzpicture}%
\begin{axis}[%
    axis lines = left,
    xmin = 0, xmax = 5,
    ymin = 0, ymax = 4,
    xlabel = {\Huge $p$}, ylabel = {\Huge $q$},
    xtick={1,2,3,4}, ytick={1,2,3},
    xmajorgrids=true, ymajorgrids=true,
]
\addplot [only marks, mark=x, mark size=6pt] table {
3   1
3   2
3   3
4   1
4   2
4   3
};
\node at (axis cs:3.5,3.5) [anchor=east] {\LARGE $r=p-2$};
\end{axis}
\end{tikzpicture} 
\\
\begin{tikzpicture}%
\begin{axis}[%
    axis lines = left,
    xmin = 0, xmax = 5,
    ymin = 0, ymax = 4,
    xlabel = {\Huge $p$}, ylabel = {\Huge $q$},
    xtick={1,2,3,4}, ytick={1,2,3},
    xmajorgrids=true, ymajorgrids=true,
]
\addplot [only marks, mark=x, mark size=6pt] table {
3	0
3	1
3	2
3   3
4	0
4	1
4	2
4	3
};

\node at (axis cs:3.5,3.5) [anchor=east] {\LARGE $r=p-2$};
\end{axis}
\end{tikzpicture} 
&
\begin{tikzpicture}%
\begin{axis}[%
    axis lines = left,
    xmin = 0, xmax = 5,
    ymin = 0, ymax = 4,
    xlabel = {\Huge $p$}, ylabel = {\Huge $q$},
    xtick={1,2,3,4}, ytick={1,2,3},
    xmajorgrids=true, ymajorgrids=true,
]
\addplot [only marks, mark=x, mark size=6pt] table {
3	0
3	1
3	2
3   3
4	0
4	1
4	2
4	3
};

\node at (axis cs:3.5,3.5) [anchor=east] {\LARGE $r=p-2$};
\end{axis}
\end{tikzpicture}
&
\begin{tikzpicture}
\begin{axis}[%
    axis lines = left,
    xmin = 0, xmax = 4,
    ymin = 0, ymax = 4,
    xlabel = {\Huge $p$}, ylabel = {\Huge $q$},
    xtick={1,2,3}, ytick={1,2,3},
    xmajorgrids=true, ymajorgrids=true,
]
\addplot [only marks, mark=x, mark size=6pt] table {
0	0
0	1
0	2
1	0
1	1
1	2
2   0 
2   1
2   2
};
\node at (axis cs:3.5,3.5) [anchor=east] {\LARGE $r=p+4$};
\end{axis}
\end{tikzpicture}
&
\begin{tikzpicture}
\begin{axis}[%
    axis lines = left,
    xmin = 0, xmax = 4,
    ymin = 0, ymax = 4,
    xlabel = {\Huge $p$}, ylabel = {\Huge $q$},
    xtick={1,2,3}, ytick={1,2,3},
    xmajorgrids=true, ymajorgrids=true,
]
\addplot [only marks, mark=x, mark size=6pt] table {
0	1
0	2
0	3
1	1
1	2
1	3
2   1
2   2
2   3
3   1
3   2
3   3
};
\node at (axis cs:3.5,3.5) [anchor=east] {\LARGE $r=p+4$};
\end{axis}
\end{tikzpicture}
\\
\begin{tikzpicture}
\begin{axis}[%
    axis lines = left,
    xmin = 0, xmax = 4,
    ymin = 0, ymax = 4,
    xlabel = {\Huge $p$}, ylabel = {\Huge $q$},
    xtick={1,2,3}, ytick={1,2,3},
    xmajorgrids=true, ymajorgrids=true,
]
\addplot [only marks, mark=x, mark size=6pt] table {
0	2
0	3
1	2
1	3
2   2
2   3
3   2
3   3
};
\node at (axis cs:3.5,3.5) [anchor=east] {\LARGE $r=p+4$};
\end{axis}
\end{tikzpicture}
&
\begin{tikzpicture}
\begin{axis}[%
    axis lines = left,
    xmin = 0, xmax = 6,
    ymin = 0, ymax = 4,
    xlabel = {\Huge $p$}, ylabel = {\Huge $q$},
    xtick={1,2,3,4,5}, ytick={1,2,3},
    xmajorgrids=true, ymajorgrids=true,
]
\addplot [only marks, mark=x, mark size=6pt] table {
4	0
4	1
4	2
4	3
5	0
5	1
5	2
5	3
};
\node at (axis cs:3.5,3.5) [anchor=east] {\LARGE $r=p-4$};
\end{axis}
\end{tikzpicture}
&
\begin{tikzpicture}
\begin{axis}[%
    axis lines = left,
    xmin = 0, xmax = 6,
    ymin = 0, ymax = 4,
    xlabel = {\Huge $p$}, ylabel = {\Huge $q$},
    xtick={1,2,3,4,5}, ytick={1,2,3},
    xmajorgrids=true, ymajorgrids=true,
]
\addplot [only marks, mark=x, mark size=6pt] table {
4	0
4	1
4	2
4	3
5	0
5	1
5	2
5	3
};
\node at (axis cs:3.5,3.5) [anchor=east] {\LARGE $r=p-4$};
\end{axis}
\end{tikzpicture}
&
\begin{tikzpicture}
\begin{axis}[%
    axis lines = left,
    xmin = 0, xmax = 6,
    ymin = 0, ymax = 4,
    xlabel = {\Huge $p$}, ylabel = {\Huge $q$},
    xtick={1,2,3,4,5}, ytick={1,2,3},
    xmajorgrids=true, ymajorgrids=true,
]
\addplot [only marks, mark=x, mark size=6pt] table {
4	0
4	1
4	2
4	3
5	0
5	1
5	2
5	3
};
\node at (axis cs:3.5,3.5) [anchor=east] {\LARGE $r=p-4$};
\end{axis}
\end{tikzpicture}
\end{tabular}%
\caption{$\Delta_L$ (symmetric traceless and transverse rank 2 tensors) towers. The empty circles mark modes which appear from the decomposition of representations of $SO(8)$ on the round sphere, but do not exist when derived in the conventional method. The excess modes are $(0,1;2)=({\bf 5,3})$ and $(1,0;1)=({\bf 4,2})$ .}
\label{fig:apx:lichntowers}
\end{figure}

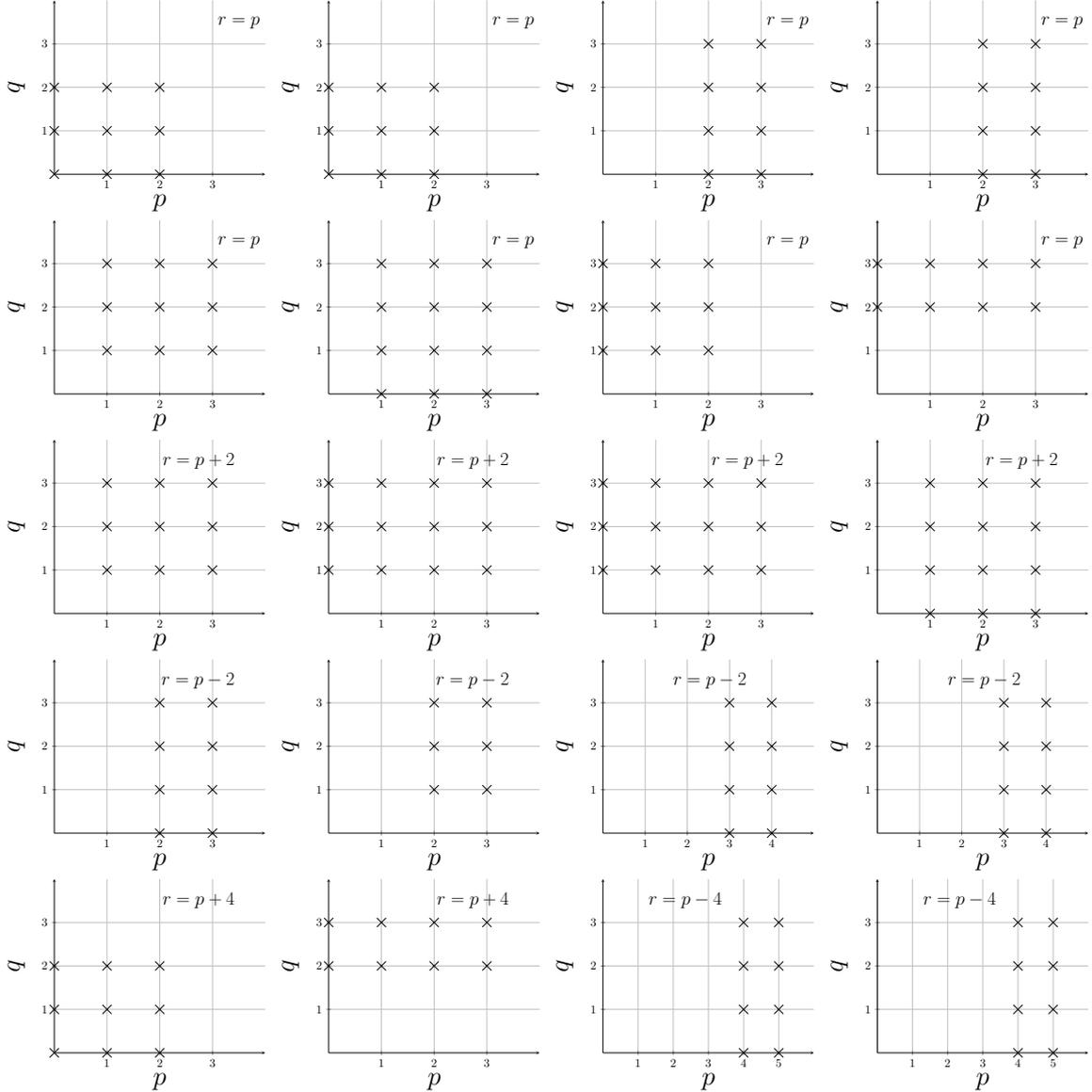
\begin{figure}[H]
\centering
\begin{tabular}{l l l l l l l l}
\begin{tikzpicture}%
\begin{axis}[%
    axis lines = left,
    xmin = 0, xmax = 4,
    ymin = 0, ymax = 4,
    xlabel = {\Huge $p$}, ylabel = {\Huge $q$},
    xtick={1,2,3}, ytick={1,2,3},
    xmajorgrids=true, ymajorgrids=true,
]
\addplot [only marks, mark=x, mark size=6pt] table {
0	0
0	1
0	2
1	0
1	1
1	2
2   0
2   1
2   2
};
\node at (axis cs:3.5,3.5) [anchor=center] {\LARGE $r=p$};
\end{axis}
\end{tikzpicture} 
&
\begin{tikzpicture}%
\begin{axis}[%
    axis lines = left,
    xmin = 0, xmax = 4,
    ymin = 0, ymax = 4,
    xlabel = {\Huge $p$}, ylabel = {\Huge $q$},
    xtick={1,2,3}, ytick={1,2,3},
    xmajorgrids=true, ymajorgrids=true,
]
\addplot [only marks, mark=x, mark size=6pt] table {
0	0
0	1
0	2
1	0
1	1
1	2
2   0
2   1
2   2
};
\node at (axis cs:3.5,3.5) [anchor=center] {\LARGE $r=p$};
\end{axis}
\end{tikzpicture}  
&
\begin{tikzpicture}%
\begin{axis}[%
    axis lines = left,
    xmin = 0, xmax = 4,
    ymin = 0, ymax = 4,
    xlabel = {\Huge $p$}, ylabel = {\Huge $q$},
    xtick={1,2,3}, ytick={1,2,3},
    xmajorgrids=true, ymajorgrids=true,
]
\addplot [only marks, mark=x, mark size=6pt] table {
2	0
2	1
2	2
2   3
3	0
3	1
3	2
3	3
};
\node at (axis cs:3.5,3.5) [anchor=center] {\LARGE $r=p$};
\end{axis}
\end{tikzpicture} 
&
\begin{tikzpicture}%
\begin{axis}[%
    axis lines = left,
    xmin = 0, xmax = 4,
    ymin = 0, ymax = 4,
    xlabel = {\Huge $p$}, ylabel = {\Huge $q$},
    xtick={1,2,3}, ytick={1,2,3},
    xmajorgrids=true, ymajorgrids=true,
]
\addplot [only marks, mark=x, mark size=6pt] table {
2	0
2	1
2	2
2   3
3	0
3	1
3	2
3	3
};
\node at (axis cs:3.5,3.5) [anchor=center] {\LARGE $r=p$};
\end{axis}
\end{tikzpicture} 
\\
\begin{tikzpicture}
\begin{axis}[%
    axis lines = left,
    xmin = 0, xmax = 4,
    ymin = 0, ymax = 4,
    xlabel = {\Huge $p$}, ylabel = {\Huge $q$},
    xtick={1,2,3}, ytick={1,2,3},
    xmajorgrids=true, ymajorgrids=true,
]
\addplot [only marks, mark=x, mark size=6pt] table {
1	1
1	2
1	3
2	1
2	2
2	3
3   1
3   2
3   3
};
\node at (axis cs:3.5,3.5) [anchor=center] {\LARGE $r=p$};
\end{axis}
\end{tikzpicture}
&
\begin{tikzpicture}
\begin{axis}[%
    axis lines = left,
    xmin = 0, xmax = 4,
    ymin = 0, ymax = 4,
    xlabel = {\Huge $p$}, ylabel = {\Huge $q$},
    xtick={1,2,3}, ytick={1,2,3},
    xmajorgrids=true, ymajorgrids=true,
]
\addplot [only marks, mark=x, mark size=6pt] table {
1	0
1	1
1	2
1	3
2	0
2	1
2	2
2	3
3   0
3   1
3   2
3   3
};
\node at (axis cs:3.5,3.5) [anchor=center] {\LARGE $r=p$};
\end{axis}
\end{tikzpicture}
&
\begin{tikzpicture}
\begin{axis}[%
    axis lines = left,
    xmin = 0, xmax = 4,
    ymin = 0, ymax = 4,
    xlabel = {\Huge $p$}, ylabel = {\Huge $q$},
    xtick={1,2,3}, ytick={1,2,3},
    xmajorgrids=true, ymajorgrids=true,
]
\addplot [only marks, mark=x, mark size=6pt] table {
0	1
0	2
0	3
1	1
1	2
1	3
2   1
2   2
2   3
};
\node at (axis cs:3.5,3.5) [anchor=center] {\LARGE $r=p$};
\end{axis}
\end{tikzpicture}
&
\begin{tikzpicture}
\begin{axis}[%
    axis lines = left,
    xmin = 0, xmax = 4,
    ymin = 0, ymax = 4,
    xlabel = {\Huge $p$}, ylabel = {\Huge $q$},
    xtick={1,2,3}, ytick={1,2,3},
    xmajorgrids=true, ymajorgrids=true,
]
\addplot [only marks, mark=x, mark size=6pt] table {
0	2
0	3
1	2
1	3
2	2
2	3
3	2
3	3
};
\node at (axis cs:3.5,3.5) [anchor=center] {\LARGE $r=p$};
\end{axis}
\end{tikzpicture}
\\
\begin{tikzpicture}%
\begin{axis}[%
    axis lines = left,
    xmin = 0, xmax = 4,
    ymin = 0, ymax = 4,
    xlabel = {\Huge $p$}, ylabel = {\Huge $q$},
    xtick={1,2,3}, ytick={1,2,3},
    xmajorgrids=true, ymajorgrids=true,
]
\addplot [only marks, mark=x, mark size=6pt] table {
1   1
1   2
1   3
2   1
2   2
2   3
3   1
3   2
3   3
};
\node at (axis cs:3.5,3.5) [anchor=east] {\LARGE $r=p+2$};
\end{axis}
\end{tikzpicture} 
&
\begin{tikzpicture}%
\begin{axis}[%
    axis lines = left,
    xmin = 0, xmax = 4,
    ymin = 0, ymax = 4,
    xlabel = {\Huge $p$}, ylabel = {\Huge $q$},
    xtick={1,2,3}, ytick={1,2,3},
    xmajorgrids=true, ymajorgrids=true,
]
\addplot [only marks, mark=x, mark size=6pt] table {
0	1
0	2
0	3
1	1
1	2
1	3
2   1
2   2
2   3
3   1
3   2
3   3
};
\node at (axis cs:3.5,3.5) [anchor=east] {\LARGE $r=p+2$};
\end{axis}
\end{tikzpicture} 
&
\begin{tikzpicture}%
\begin{axis}[%
    axis lines = left,
    xmin = 0, xmax = 4,
    ymin = 0, ymax = 4,
    xlabel = {\Huge $p$}, ylabel = {\Huge $q$},
    xtick={1,2,3}, ytick={1,2,3},
    xmajorgrids=true, ymajorgrids=true,
]
\addplot [only marks, mark=x, mark size=6pt] table {
0	1
0	2
0	3
1	1
1	2
1	3
2   1
2   2
2   3
3   1
3   2
3   3
};
\node at (axis cs:3.5,3.5) [anchor=east] {\LARGE $r=p+2$};
\end{axis}
\end{tikzpicture} 
&
\begin{tikzpicture}%
\begin{axis}[%
    axis lines = left,
    xmin = 0, xmax = 4,
    ymin = 0, ymax = 4,
    xlabel = {\Huge $p$}, ylabel = {\Huge $q$},
    xtick={1,2,3}, ytick={1,2,3},
    xmajorgrids=true, ymajorgrids=true,
]
\addplot [only marks, mark=x, mark size=6pt] table {
1	0
1	1
1	2
1	3
2	0
2	1
2	2
2	3
3   0
3   1
3   2
3   3
};
\node at (axis cs:3.5,3.5) [anchor=east] {\LARGE $r=p+2$};
\end{axis}
\end{tikzpicture}
\\
\begin{tikzpicture}%
\begin{axis}[%
    axis lines = left,
    xmin = 0, xmax = 4,
    ymin = 0, ymax = 4,
    xlabel = {\Huge $p$}, ylabel = {\Huge $q$},
    xtick={1,2,3}, ytick={1,2,3},
    xmajorgrids=true, ymajorgrids=true,
]
\addplot [only marks, mark=x, mark size=6pt] table {
2	0
2	1
2	2
2	3
3	0
3	1
3	2
3	3
};
\node at (axis cs:3.5,3.5) [anchor=east] {\LARGE $r=p-2$};
\end{axis}
\end{tikzpicture} 
&
\begin{tikzpicture}%
\begin{axis}[%
    axis lines = left,
    xmin = 0, xmax = 4,
    ymin = 0, ymax = 4,
    xlabel = {\Huge $p$}, ylabel = {\Huge $q$},
    xtick={1,2,3}, ytick={1,2,3},
    xmajorgrids=true, ymajorgrids=true,
]
\addplot [only marks, mark=x, mark size=6pt] table {
2	1
2	2
2	3
3	1
3	2
3	3
};
\node at (axis cs:3.5,3.5) [anchor=east] {\LARGE $r=p-2$};
\end{axis}
\end{tikzpicture} 
&
\begin{tikzpicture}%
\begin{axis}[%
    axis lines = left,
    xmin = 0, xmax = 5,
    ymin = 0, ymax = 4,
    xlabel = {\Huge $p$}, ylabel = {\Huge $q$},
    xtick={1,2,3,4}, ytick={1,2,3},
    xmajorgrids=true, ymajorgrids=true,
]
\addplot [only marks, mark=x, mark size=6pt] table {
3	0
3	1
3	2
3   3
4	0
4	1
4	2
4	3
};
\node at (axis cs:3.5,3.5) [anchor=east] {\LARGE $r=p-2$};
\end{axis}
\end{tikzpicture} 
&
\begin{tikzpicture}%
\begin{axis}[%
    axis lines = left,
    xmin = 0, xmax = 5,
    ymin = 0, ymax = 4,
    xlabel = {\Huge $p$}, ylabel = {\Huge $q$},
    xtick={1,2,3,4}, ytick={1,2,3},
    xmajorgrids=true, ymajorgrids=true,
]
\addplot [only marks, mark=x, mark size=6pt] table {
3	0
3	1
3	2
3   3
4	0
4	1
4	2
4	3
};
\node at (axis cs:3.5,3.5) [anchor=east] {\LARGE $r=p-2$};
\end{axis}
\end{tikzpicture}
\\
\begin{tikzpicture}
\begin{axis}[%
    axis lines = left,
    xmin = 0, xmax = 4,
    ymin = 0, ymax = 4,
    xlabel = {\Huge $p$}, ylabel = {\Huge $q$},
    xtick={1,2,3}, ytick={1,2,3},
    xmajorgrids=true, ymajorgrids=true,
]
\addplot [only marks, mark=x, mark size=6pt] table {
0	0
0	1
0	2
1	0
1	1
1	2
2   0 
2   1
2   2
};
\node at (axis cs:3.5,3.5) [anchor=east] {\LARGE $r=p+4$};
\end{axis}
\end{tikzpicture}
&
\begin{tikzpicture}
\begin{axis}[%
    axis lines = left,
    xmin = 0, xmax = 4,
    ymin = 0, ymax = 4,
    xlabel = {\Huge $p$}, ylabel = {\Huge $q$},
    xtick={1,2,3}, ytick={1,2,3},
    xmajorgrids=true, ymajorgrids=true,
]
\addplot [only marks, mark=x, mark size=6pt] table {
0	2
0	3
1	2
1	3
2   2
2   3
3   2
3   3
};
\node at (axis cs:3.5,3.5) [anchor=east] {\LARGE $r=p+4$};
\end{axis}
\end{tikzpicture}
&
\begin{tikzpicture}
\begin{axis}[%
    axis lines = left,
    xmin = 0, xmax = 6,
    ymin = 0, ymax = 4,
    xlabel = {\Huge $p$}, ylabel = {\Huge $q$},
    xtick={1,2,3,4,5}, ytick={1,2,3},
    xmajorgrids=true, ymajorgrids=true,
]
\addplot [only marks, mark=x, mark size=6pt] table {
4	0
4	1
4	2
4	3
5	0
5	1
5	2
5	3
};
\node at (axis cs:3.5,3.5) [anchor=east] {\LARGE $r=p-4$};
\end{axis}
\end{tikzpicture}
&
\begin{tikzpicture}
\begin{axis}[%
    axis lines = left,
    xmin = 0, xmax = 6,
    ymin = 0, ymax = 4,
    xlabel = {\Huge $p$}, ylabel = {\Huge $q$},
    xtick={1,2,3,4,5}, ytick={1,2,3},
    xmajorgrids=true, ymajorgrids=true,
]
\addplot [only marks, mark=x, mark size=6pt] table {
4	0
4	1
4	2
4	3
5	0
5	1
5	2
5	3
};
\node at (axis cs:3.5,3.5) [anchor=east] {\LARGE $r=p-4$};
\end{axis}
\end{tikzpicture}
\end{tabular}%
\caption{$Q$ towers.}
\label{fig:apx:Qtowers}
\end{figure}

%
%
%
%



\end{document}